\documentclass[usegraphicx]{emulateapj}

\def\largeplottwo#1#2{\centering \leavevmode
\includegraphics[width=130mm]{#1}
 \hfil \includegraphics[width=130mm]{#2} }

\def\plottwo#1#2{\centering \leavevmode
\includegraphics[width=120mm]{#1}
 \hfil \includegraphics[width=120mm]{#2} }

\def\plotsix#1#2#3#4#5#6{\centering \leavevmode
\includegraphics[width=85mm]{#1} \includegraphics[width=85mm]{#2}
\hfil \includegraphics[width=85mm]{#3} \includegraphics[width=85mm]{#4}
\hfil \includegraphics[width=85mm]{#5} \includegraphics[width=85mm]{#6}}

\begin{document}
\title{Infrared Two-Color Diagrams for AGB stars, post-AGB stars, and Planetary Nebulae}

\author{Kyung-Won Suh\altaffilmark{1}}

\altaffiltext{1}{Department of Astronomy and Space Science, Chungbuk National University,
        Cheongju-City, 362-763, Republic of Korea; kwsuh@chungbuk.ac.kr}

\slugcomment{submitted to the Astrophysical Journal 2015 April 26; accepted
2015 June 6}

\begin{abstract}

We present various infrared two-color diagrams (2CDs) for AGB stars, post-AGB
stars, and Planetary Nebulae (PNe) and investigate possible evolutionary
tracks. We use catalogs from the available literature for the sample of 4903
AGB stars (3373 O-rich; 1168 C-rich; 362 S-type), 660 post-AGB stars (326
post-AGB; 334 pre-PNe), and 1510 PNe in our Galaxy. For each object in the
catalog, we cross-identify the $IRAS$, $AKARI$, $MSX$, and 2MASS
counterparts. The IR 2CDs can provide useful information about the structure
and evolution of the dust envelopes as well as the central stars. To find
possible evolutionary tracks from AGB stars to PNe on the 2CDs, we
investigate spectral evolution of post-AGB stars by making simple but
reasonable assumptions on the evolution of the central star and dust shell.
We perform radiative transfer model calculations for the detached dust shells
around evolving central stars in the post-AGB phase. We find that the
theoretical dust shell model tracks using dust opacity functions of amorphous
silicate and amorphous carbon roughly coincide with the densely populated
observed points of AGB stars, post-AGB stars, and PNe on various IR 2CDs.
Even though some discrepancies are inevitable, the end points of the
theoretical post-AGB model tracks are generally converged to the region of
the observed points of PNe on most 2CDs.

\end{abstract}


\section{Introduction}

Asymptotic giant branch (AGB) stars are generally classified as O-rich (M-type)
or C-rich (C-type) based on the chemistry of the photosphere and/or the outer
envelope. When an AGB star of intermediate mass range goes through C dredge-up
process and thus the C/O ratio is larger than 1, O-rich dust grain formation
ceases and the star may become a carbon star (e.g., Iben 1981; Chan \& Kwok
1990). S stars are generally regarded as intermediate between M-type and carbon
stars in their properties (Lloyd Evans \& Little-Marenin 1999). However, this
M-S-C evolutionary sequence can be different depending on the mass and
metallicity (e.g., Groenewegen et al. 1995; Suh 2014). Dust envelopes around
AGB stars are believed to be a main source of interstellar dust. The outflowing
envelopes around AGB stars are very suitable places for massive dust formation.
Nearly all AGB stars can be identified as Long-Period Variables (LPVs). The AGB
phase of the LPV is characterized by dusty stellar winds with high mass-loss
rates ($10^{-8} - 10^{-4} M_{\odot}/yr$; e.g., Loup et al. 1993).

As the star leaves the AGB phase, its mass-loss rate decreases significantly
and the star may become hot enough to ionize its circumstellar material. When
mass-loss reduces the mass of the remaining H-rich envelope below $\sim$
$10^{-2} M_{\odot}$ (the exact value depends on the initial mass), the stellar
envelope begins to shrink and the effective temperature starts to increase
(e.g., Sch\"{o}nberner 1983) until the central star is hot enough ($\sim$ 30000
K) to ionize the circumstellar nebula. If the temperature increases on a
timescale shorter than the dispersion time of the matter previously ejected by
the star, a planetary nebula (PN) will appear. Planetary nebulae (PNe) are
believed to be a common end-point of stellar evolution for a large fraction of
all stars between 1 and 10 $M_{\odot}$.

The intermediate phase between the end of the AGB phase and the PN phase is
called the post-AGB phase. During the the post-AGB phase, the dust shell formed
in the AGB phase detaches from the central star and becomes optically thin
after a few hundred years (e.g., Hrivnak et al. 1989). There could be some
post-AGB objects that have left the AGB stage but will evolve to the white
dwarf stage without ever becoming a PN (e.g., Zuckerman 1978; Szczerba et al.
2007). The evolutionary lifetime of a PN is critically dependent on the
core-mass of its central star (e.g., Renzini 1981). A low-mass star evolves too
slowly to become a PN and a high-mass star is luminous too briefly to be
detected. Therefore, only stars with core-masses in a narrow mass range ($\sim$
0.6 $M_{\sun}$) would be seen as PNe (e.g., Kwok 2000).

The $Infrared$ $Astronomical$ $Satellite$ ($IRAS$) Point Source Catalog (PSC)
(version 2.1) provided useful photometric data in four bands (12, 25, 60, and
100 $\mu$m). In characterizing the circumstellar environments of AGB and
post-AGB stars, the two-color diagram (2CD) in the $IRAS$ PSC has been useful
(e.g., Suh \& Kwon 2011). The $IRAS$ Low Resolution Spectrograph (LRS;
$\lambda$ = 8$-$22 $\mu$m) data are useful to identify important features of
dust grains.  Kwok et al. (1997) used $IRAS$ LRS to identify the class E (the
10 $\mu$m silicate feature in emission), class A (the 10 $\mu$m silicate
feature in absorption), class C (the 11 $\mu$m SiC dust emission), and class P
(the 11.3 $\mu$m or 12.5 $\mu$m emission features that are attributed to
polycyclic aromatic hydrocarbon; PAH).

The Midcourse Space Experiment ($MSX$; Egan et al. 2003) surveyed the Galactic
plane in four bands (8.28, 12.13, 14.65, and 21.34 $\mu$m). $AKARI$ (Murakami
et al. 2007) provided PSC data in two bands (9 and 18 $\mu$m) from an all-sky
survey. The PSC from the two micron all sky survey (2MASS; Cutri et al. 2003)
contains accurate positions and fluxes in $J$ (1.25 $\mu$m), $H$ (1.65 $\mu$m),
and $K_s$ (2.17 $\mu$m) bands.

In this paper, we present various IR 2CDs for AGB stars, post-AGB stars, and
PNe and investigate possible evolutionary tracks. We use catalogs from the
available literature for the sample stars in our Galaxy. For each object in
the catalog, we cross-identify the $IRAS$, $AKARI$, $MSX$, and 2MASS
counterparts. To investigate possible evolutionary tracks from AGB stars to
PNe on the 2CDs, we make simple but reasonable theoretical models on the
evolution of the central star and dust shell in the post-AGB phase. We
perform radiative transfer model calculations for detached dust shells around
evolving central stars in the post-AGB phase. We compare the theoretical
model tracks with observations of AGB stars, post-AGB stars, and PNe and
discuss possible evolutionary tracks on the IR 2CDs.

\begin{table*}
\centering
\caption{Sample of AGB stars, post-AGB stars, and PNe}
\begin{tabular}{lllllllllll}
\hline
\hline
Class  &Reference & Number & $IRAS$ PSC & $AKRAI$ PSC & $MSX$ PSC &2MASS & A$^*$  & E$^*$ & C$^*$ & P$^*$ \\
\hline
O-AGB & Kwon \& Suh (2012) & 3373  & 3373  & 2708 & 1753 &2708 &186 &1125 &0 &44       \\
C-AGB & Suh \& Kwon (2011) & 1168  &  1168 & 1012 & 687 & 1012 &0  &0 &713 &11       \\
S-AGB & Suh \& Kwon (2011) & 362  & 362   &  336 & 128 &336 & 0 &29 &1 &0      \\
post-AGB & Szczerba et al. (2007) & 326  & 236 & 254 & 117 &318 &4 &14 &1 &3  \\
Pre-PNe  & Kohoutek (2001) & 334   & 326 & 253 & 177 & 253 &15 &24 &6 &4   \\
PNe  & Kohoutek (2001) & 1510   &  927 & 808 & 414  &808 &1 &15 &0 &13       \\
\hline
\end{tabular}
\begin{flushleft}
$^*$: The $IRAS$ LRS class.
\end{flushleft}
\end{table*}

\section{Sample stars and IR Two-Color Diagrams}

We use catalogs from the available literature for the sample of 4903 AGB stars
(3373 O-AGB; 1168 C-AGB; 362 S-type), 660 post-AGB stars (326 post-AGB; 334
pre-PN), and 1510 PNe. For each object in the catalog, we cross-identify the
$IRAS$ PSC, $AKARI$ PSC, $MSX$ PSC, and 2MASS counterparts as we will explain
in subsections.

In Table 1, we list the reference, total number of objects, and numbers of the
cross-identified $IRAS$ PSC, $AKARI$ PSC, $MSX$ PSC, and 2MASS counterparts. We
also list the number of sources according to the $IRAS$ LRS classification.
Class E and A objects show O-rich dust features and class C and P objects show
C-rich dust features (see section 1).

There are considerable differences in the angular resolutions of different data
sets. The angular resolutions for the $IRAS$, $MSX$, $AKARI$, and 2MASS are
0.75$\times$4.5-4.6$\arcmin$, 18.3$\arcsec$, 9.4$\arcsec$ (in 9 $\mu$m band),
and 2$\arcsec$, respectively. The differing spatial resolutions may have
impacts on the photometry used in the 2CDs. The 2MASS flux presumably includes
only the central star, while the $IRAS$ fluxes may include the extended outer
envelope.

Because $IRAS$ has very low angular resolution, the survey regions may suffer
from confusion problems. So the 2MASS counterpart obtained from the $IRAS$ PSC
position can be a different object. The 2MASS counterpart obtained from the
position data from $AKARI$ which has a higher resolution would be much more
reliable.

\subsection{AGB stars}

Suh \& Kwon (2011) presented a catalog of AGB stars for 3003 O-rich, 1168
C-rich, and 362 S-type objects in our Galaxy. For the catalog, they compiled
previous works with verifying processes from the sources listed in the $IRAS$
PSC. Kwon \& Suh (2012) presented a revised sample of 3373 O-rich AGB stars.

By using the position information in version 2.1 of the $IRAS$ PSC, we
cross-identify the $AKARI$ and $MSX$ counterparts by finding the nearest source
within 30$\arcsec$ for each object. For 2MASS, we find the closest counterpart
in the position within 30$\arcsec$ using the position information of the
cross-identified $AKARI$ PSC source.

\begin{table}
\centering
\caption{Duplicate classification of AGB and post-AGB stars}
\begin{tabular}{lllll}
\hline
\hline
         & O-AGB & C-AGB &  post-AGB  & Pre-PNe  \\
\hline
O-AGB    & \textbf{3373}  & 0  & 34  &74    \\
C-AGB    & 0  & \textbf{1168}  & 12  &  23   \\
post-AGB & 34    & 12  &  \textbf{326}  &  100        \\
Pre-PNe  & 74   & 24  & 100  & \textbf{334}       \\
\hline
\end{tabular}
\end{table}

\subsection{Post-AGB stars}

The intermediate phase between the end of the AGB phase and the PN phase is
called the post-AGB phase (e.g., Szczerba et al. 2007). It was formerly also
referred to as the proto-PN phase or the pre-PN phase (e.g., Zuckerman 1978;
Parthasarathy \& Pottasch 1986; Hrivnak et al. 1989; Kwok 2000; Kohoutek 2001).

For this paper, we use the list of 326 `very likely post-AGB objects' from the
catalog of post-AGB stars by Szczerba et al. (2007) and 334 pre-PNe from the
catalogue of Galactic Planetary Nebulae (Updated Version 2000; Kohoutek 2001).
We regard the 660 objects in both catalogs (post-AGB objects and pre-PNe) as
post-AGB stars.

Because AGB stars and post-AGB stars share some similarities, they could have
been duplicately classified. Table 2 lists them. For S type AGB stars, there
are no duplicated objects.

For post-AGB stars, we use the $IRAS$ PSC, $MSX$, and 2MASS counterparts listed
in Szczerba et al. (2007). We cross-identify the $AKARI$ counterparts by
finding the nearest source within 30$\arcsec$ using the position information in
Szczerba et al. (2007) for each object.

For pre-PNe, we use the $IRAS$ PSC counterparts listed in Kohoutek (2001). We
cross-identify the $AKARI$ and $MSX$ counterparts by finding the nearest source
within 30$\arcsec$ using the position information in Kohoutek (2001). For the
2MASS counterpart, we find the closest source in the position within
30$\arcsec$ using the position information of the cross-identified $AKARI$ PSC
source.

\subsection{Planetary Nebulae}

For this paper, we use the catalogue of Galactic Planetary Nebulae (Updated
Version 2000; Kohoutek 2001). This catalogue contains 1510 objects classified
as galactic PNe up to the end of 1999. Valuable information about the objects
including their identification charts and the lists of references can be found
in the Strasbourg-ESO Catalogue of Galactic Planetary Nebulae (Acker et al.
1992; SECGPN) and in its First Supplement (Acker et al. 1996). Kerber et al.
(2003) presented more accurate position data for 1312 PNe.

There have been significant new discoveries of PNe after Kohoutek (2001). The
most important contribution would be the Macquarie/AAO/Strasbourg H$\alpha$
(MASH) catalogue (Parker et al. 2006) which represents the largest
incremental increase in Galactic PN numbers to date. However, MASH PNe are
typically more evolved, obscured, and of lower surface brightness than those
found in most previous surveys. Therefore, we do not use the newly discovered
PNe because they are not very useful for the purpose of this work.

For PNe, we use $IRAS$ PSC counterpart listed in Kohoutek (2001). We
cross-identify the $AKARI$ and $MSX$ counterparts by finding the nearest source
within 30$\arcsec$ using the position information in Kohoutek (2001) and Kerber
et al. (2003). For the 2MASS counterpart, we find the closest source in the
position within 30$\arcsec$ using the position information of the
cross-identified $AKARI$ PSC source.

\subsection{IR Two-Color Diagrams}

Infrared fluxes from the $IRAS$ (PSC; 12, 25, 60, and 100 $\mu$m), $MSX$ (PSC;
8.28, 12.13, 14.65, and 21.34 $\mu$m), $AKARI$ (PSC; 9 and 18 $\mu$m) and 2MASS
(1.25, 1.65, and 2.17 $\mu$m) are available for the sample of AGB stars,
post-AGB stars, and PNe as discussed in previous subsections. The large number
of observations can be used to form various 2CDs that can be compared with
theoretical model predictions.

The color index is defined by
\begin{eqnarray}
  M_{\lambda 1} - M_{\lambda 2} &=& 2.5 \log_{10} {{F_{\lambda 2} / ZMC_{\lambda 2}} \over {F_{\lambda 1} / ZMC_{\lambda 1}}}
\end{eqnarray}
\noindent where $ZMC_{\lambda i}$ means the zero magnitude calibration at
given wavelength ($\lambda i$) (see Suh \& Kwon 2011 for details).

Figures 1, 2, 3, and 4 show $IRAS$, NIR, $AKARI$, and $MSX$ 2CDs for AGB stars,
post-AGB stars, and PNe compared with theoretical models. The small symbols are
observational data and the lines with symbols are theoretical model
calculations. For each figure, the upper panel plots all observed objects and
theoretical model tracks for AGB stars. The lower panel plots observed objects
of post-AGB stars and PNe and theoretical model tracks for post AGB stars. We
will discuss the theoretical models in sections 3 and 4. We will compare the
theoretical models with the observations on the IR 2CDs in section 5.

For AGB stars, we plot only those objects with good quality (q = 3) at any
wavelength on all 2CDs using the $IRAS$, $AKARI$, 2MASS, and $MSX$ data. On the
other hand, we plot all objects for post-AGB stars and PNe. For the post-AGB
stars and PNe with good quality (q = 3) data, we plot them by thicker symbols.
For the post-AGB stars and PNe which have the $IRAS$ LRS data, we indicate the
$IRAS$ LRS class by corresponding symbols. Class E and A objects show O-rich
dust features and class C and P objects show C-rich dust features. The 42
post-AGB stars which are known to be binary systems (Szczerba et al. 2007) are
marked by the green hexagonal symbol.

For AGB stars, the objects in the upper-right regions on any 2CDs have thicker
dust shells with large optical depths. Generally, the [12]$-$[25],
[14.65]$-$[21.34], and [25]$-$[60] colors for PNe are redder than those for AGB
stars with thick dust shells. The [8.28]$-$[14.65] colors for PNe are
comparable to those for AGB stars with thick dust shells. However,
$K_{s}$$-$[12] and [9]$-$[12] colors for PNe are bluer than those for AGB stars
with thick dust shells. These effects are clearly displayed in the 2CDs. The
locations of PNe look to be converged toward upper-right regions on the two
2CDs using [25]$-$[60] versus [12]$-$[25] and [8.28]$-$[14.65] versus
[14.65]$-$[21.34]. But for the 2CD using [12]$-$[25] versus $K_{s}$$-$[12], the
convergence occurs toward the upper-middle region. For the 2CD using [9]$-$[12]
versus [12]$-$[25], the convergence occurs toward the right region.

\begin{figure*}
\centering
\largeplottwo{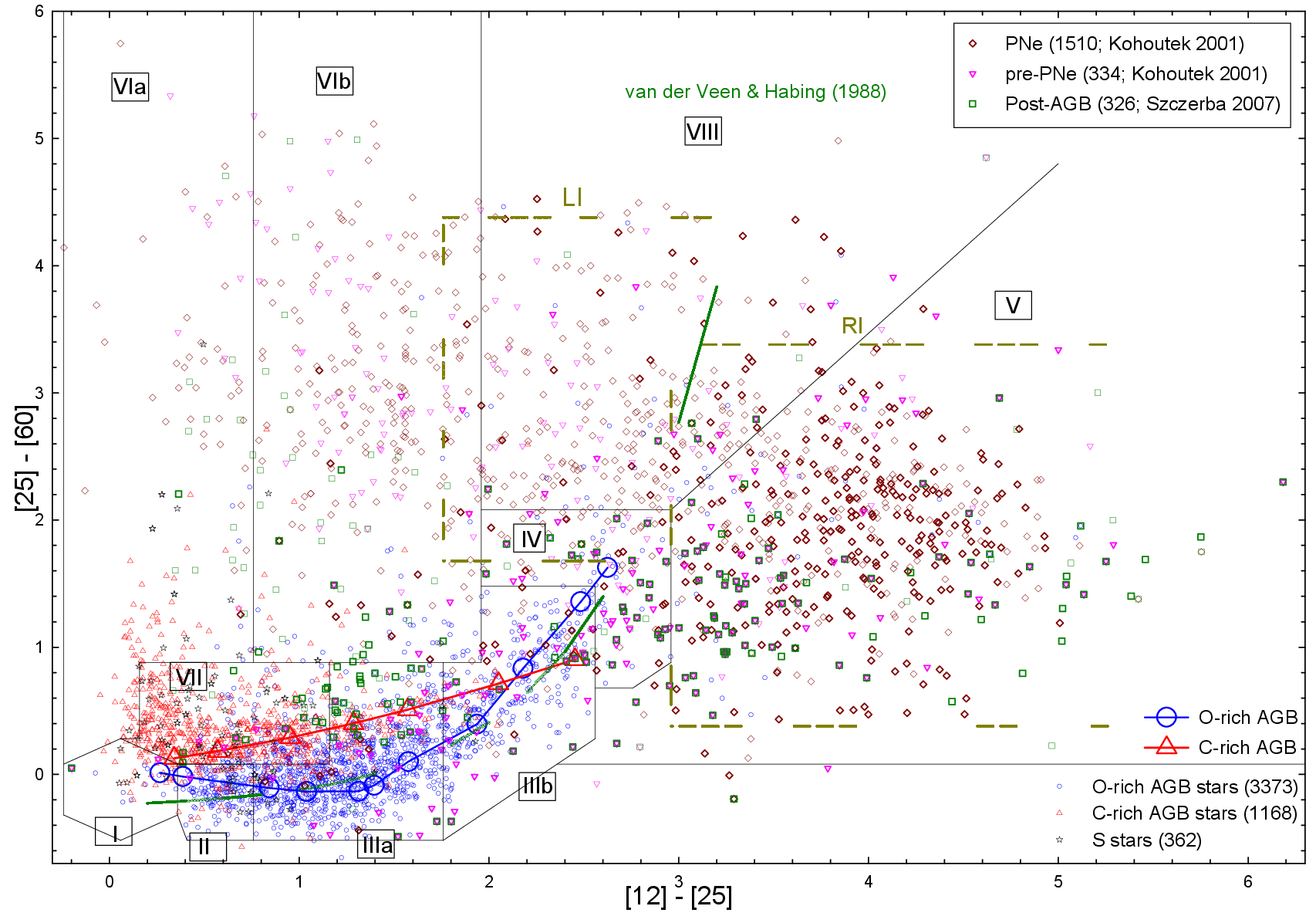}{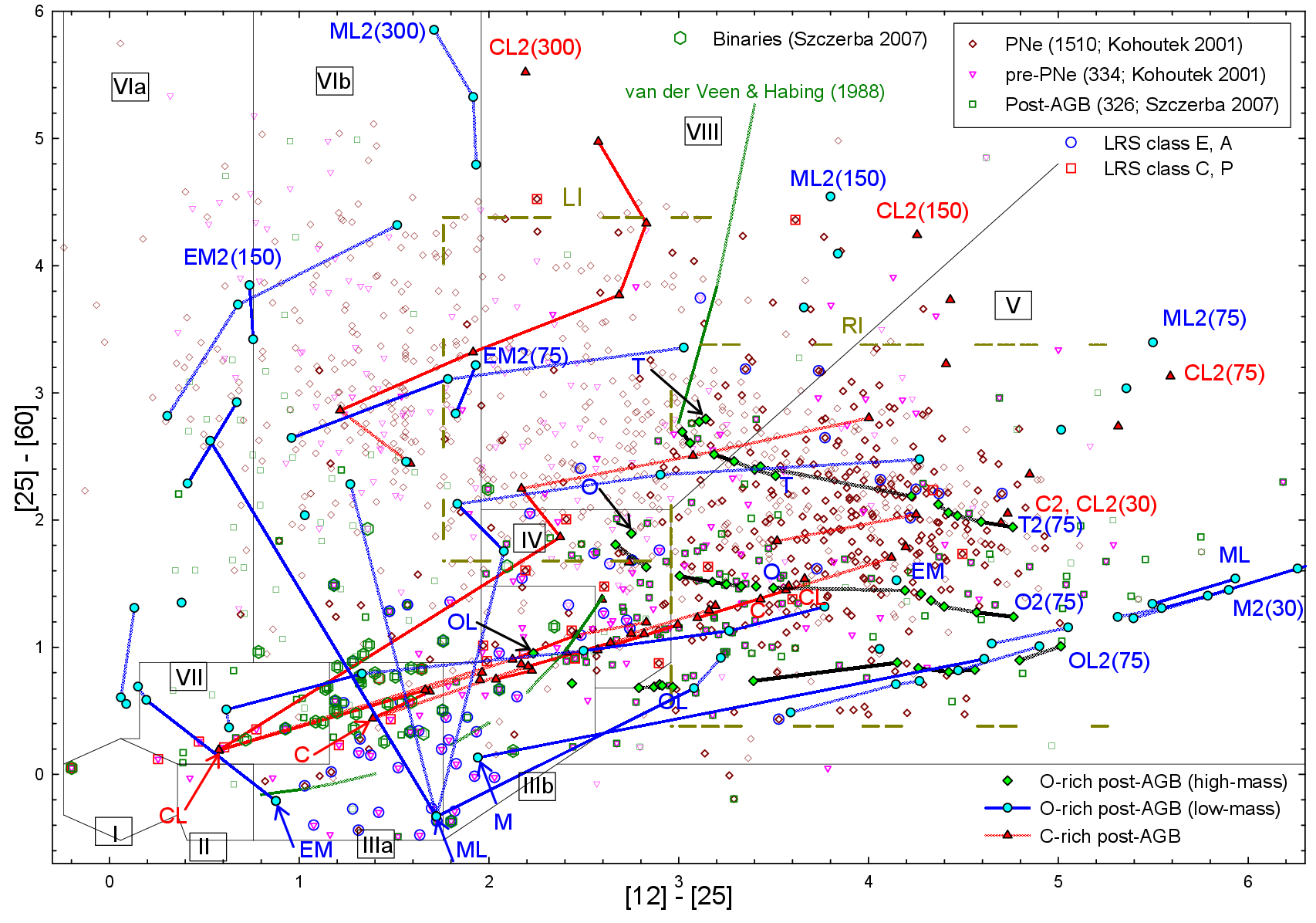}\caption{$IRAS$ 2CDs for AGB stars,
post-AGB stars, and planetary nebulae compared with theoretical models.
See sections 3 - 5 for discussion on the theoretical models.}
\end{figure*}

\begin{figure*}
\centering
\plottwo{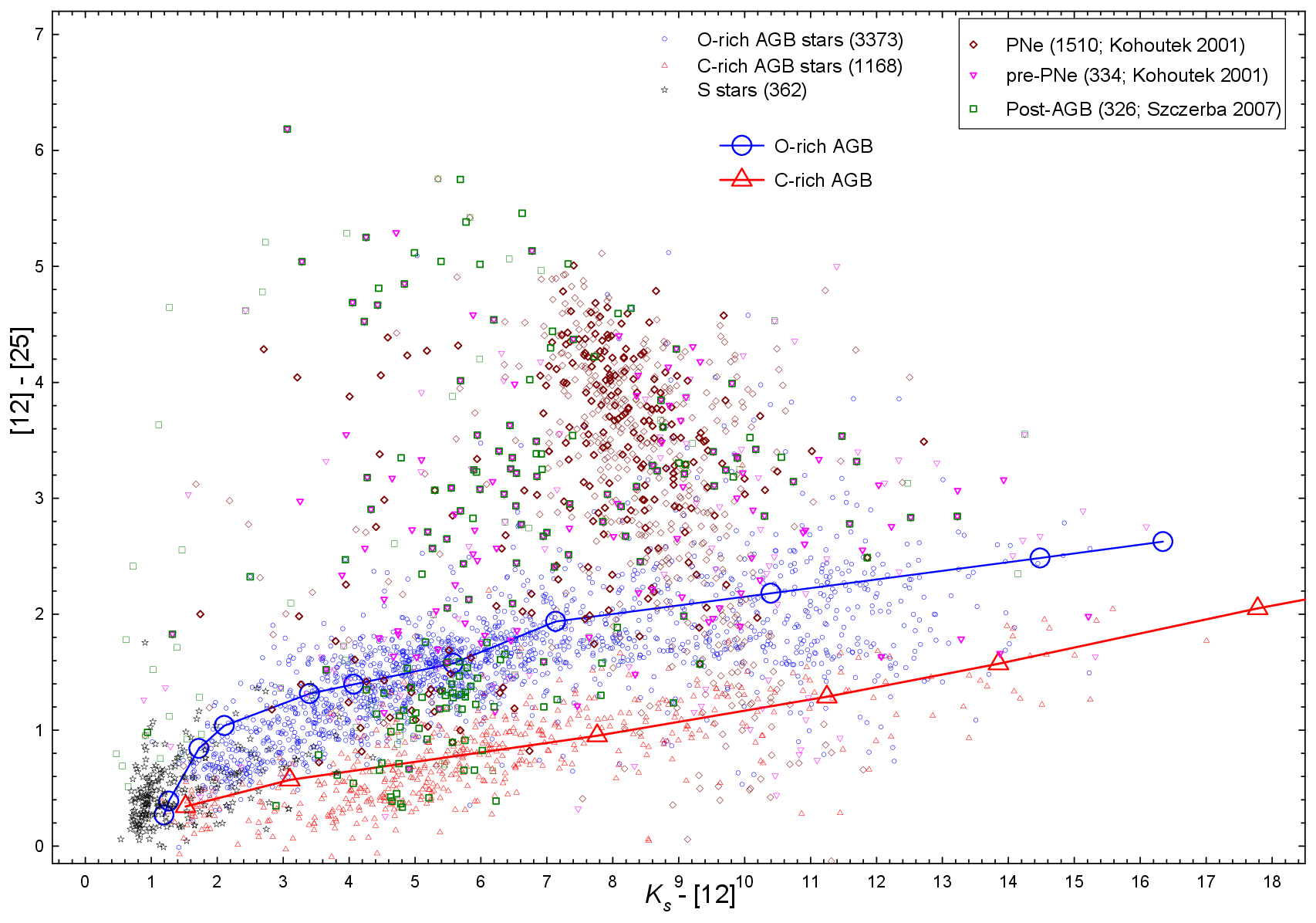}{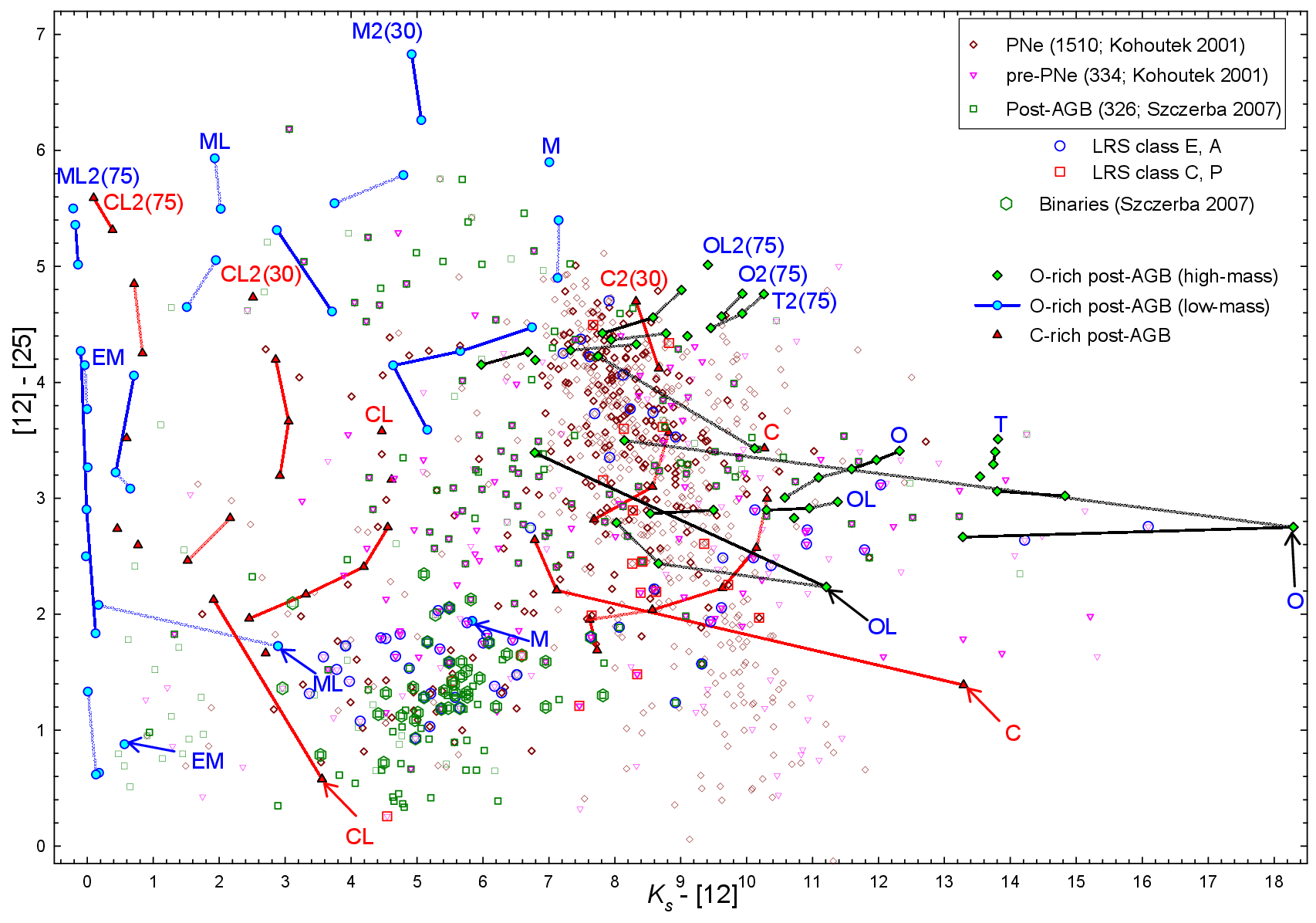}\caption{NIR-$IRAS$ 2CDs for AGB stars,
post-AGB stars, and planetary nebulae compared with theoretical models.}
\end{figure*}

\begin{figure*}
\centering
\plottwo{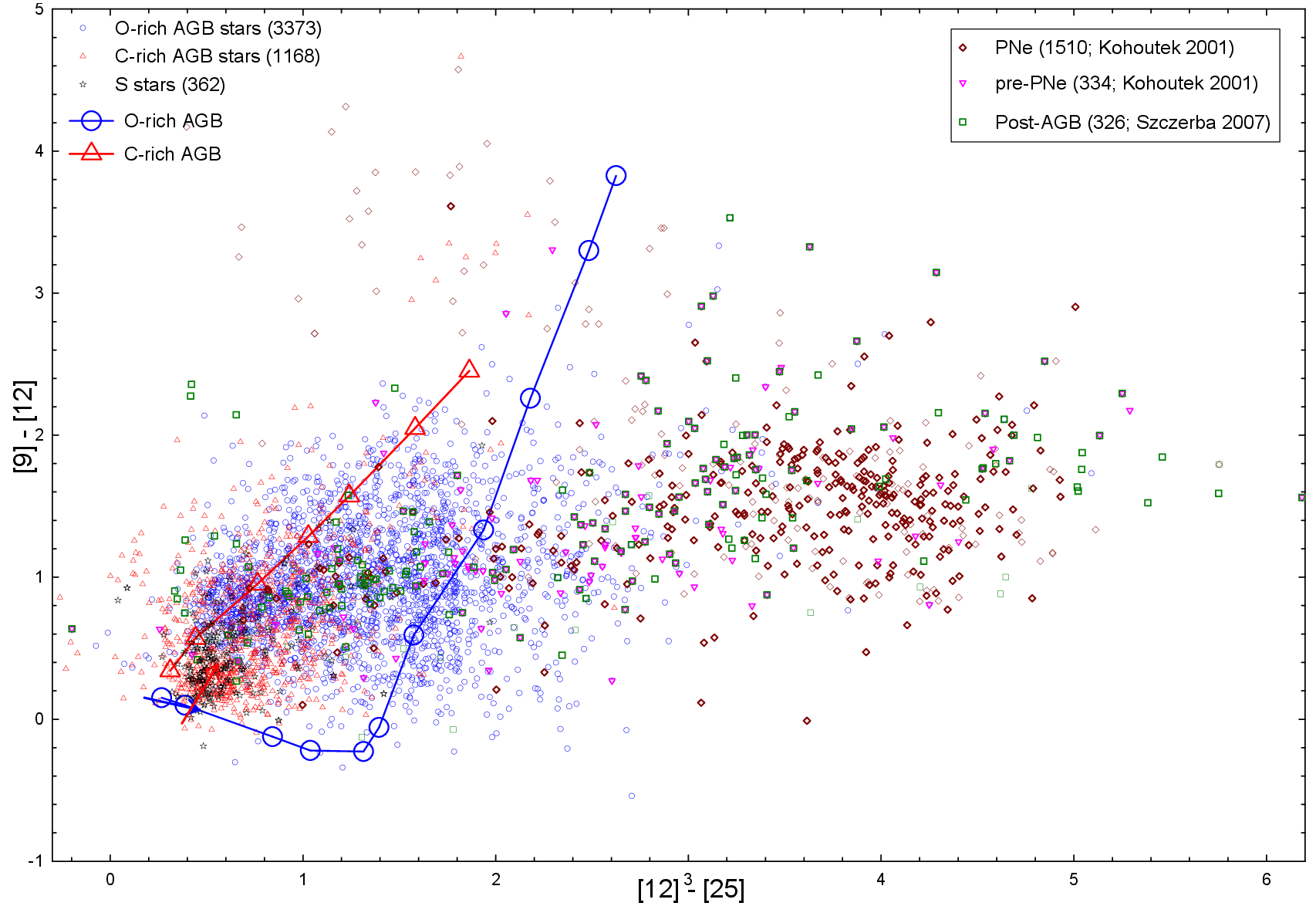}{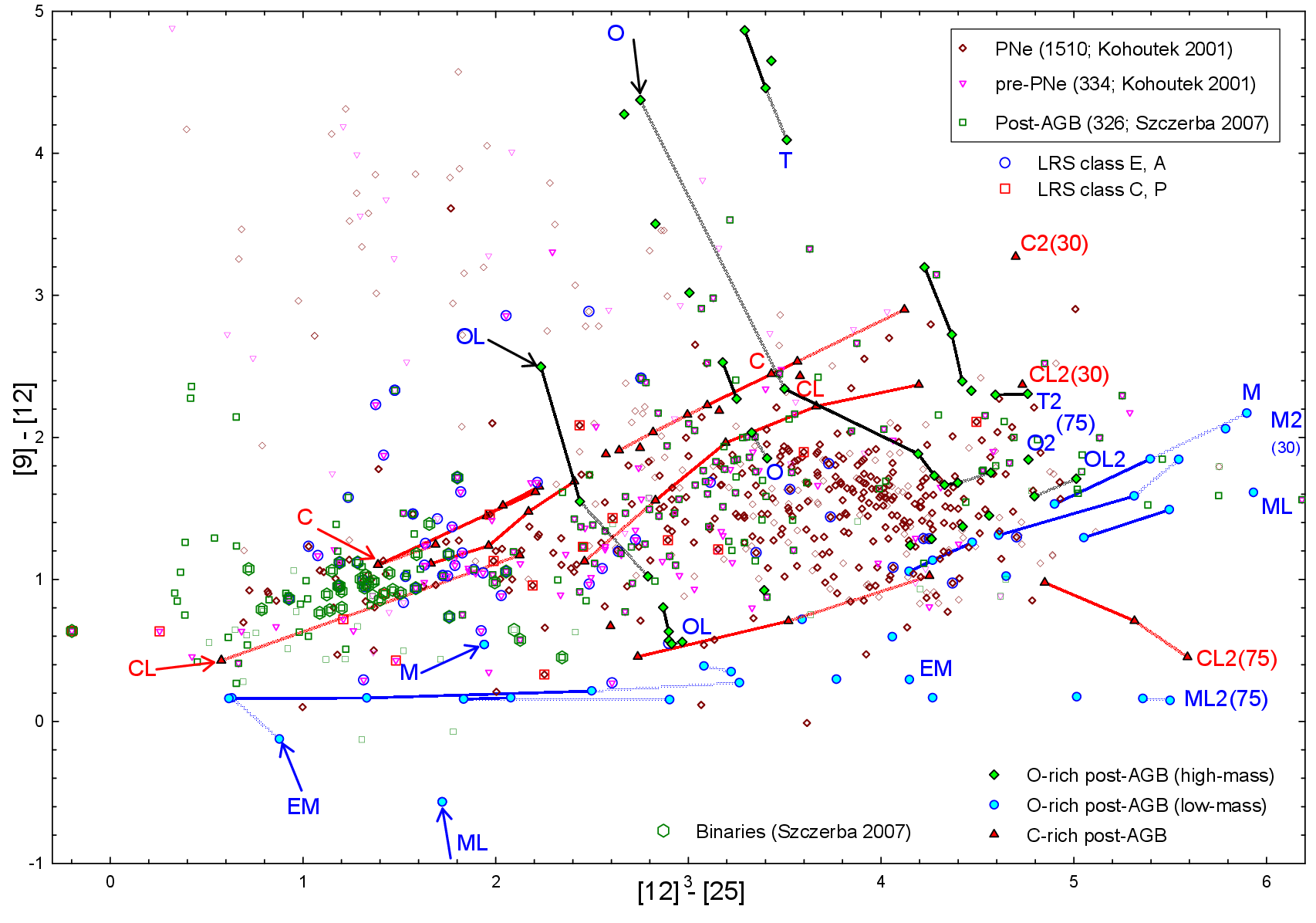}\caption{$AKARI$-$IRAS$ 2CDs for AGB stars,
post-AGB stars, and planetary nebulae compared with theoretical models.}
\end{figure*}

\begin{figure*}
\centering
\plottwo{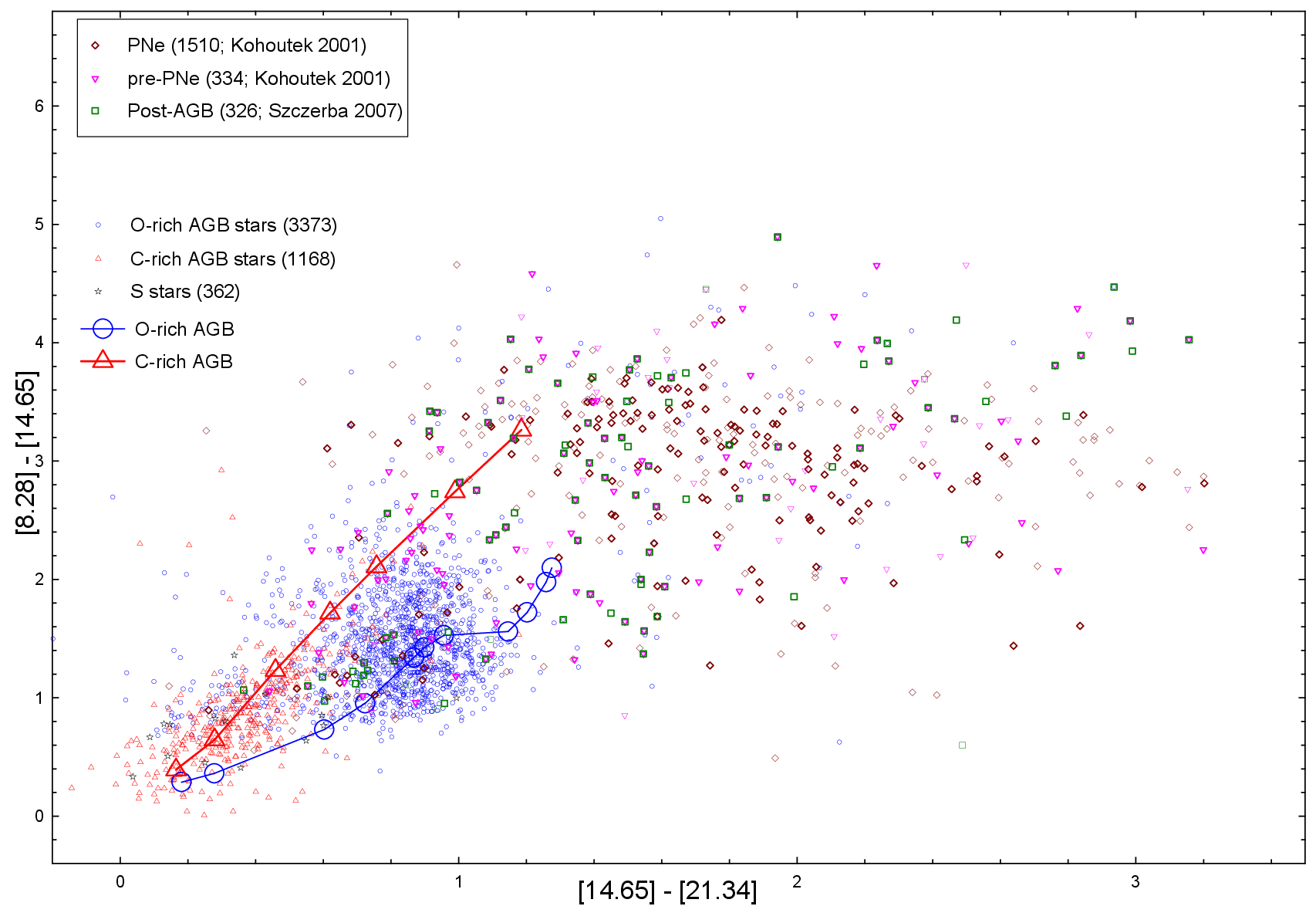}{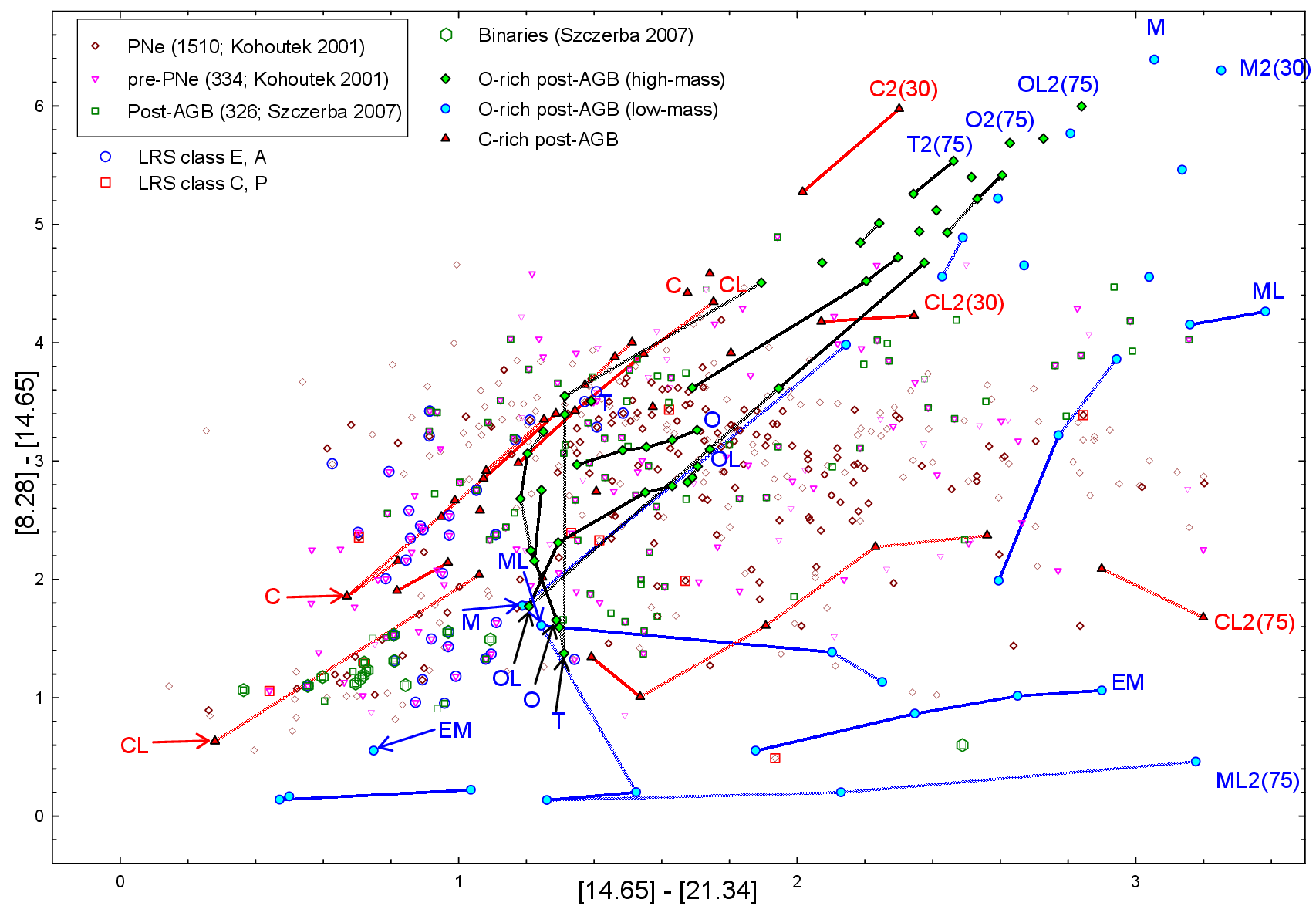}\caption{MSX 2CDs for AGB stars,
post-AGB stars, and planetary nebulae compared with theoretical models.}
\end{figure*}

\section{Theoretical dust shell models}

To investigate the spectral evolution of AGB and post-AGB stars on the IR
2CDs, we use radiative transfer models for spherically symmetric dust shells.
We assume that the central star emits blackbody radiation (see section 6.1
for a discussion on the assumptions).

\subsection{Radiative transfer model calculations}

In this work, we use the radiative transfer code RADMC-3D
(http://www.ita.uni-heidelberg.de/$\sim$dullemond/software/radmc-3d/) to
calculate the model SEDs for dust shells around AGB stars and pos-AGB stars.
RADMC-3D is based on the Monte Carlo simulation method of Bjorkman \& Wood
(2001) for investigating dust continuum radiative transfer processes. For a
spherically symmetric dust shell, we have compared the model results obtained
by the RADMC-3D code with those obtained by the radiative transfer codes
CSDUST3 (Egan et al. 1988) and DUSTY (Ivezi\'{c} \& Elitzur (1997). All of
these three codes produced essentially identical results for the same model
parameters.

We assume a spherically symmetric dust shell around a single star. We use
similar schemes for dust density distribution as those used by Suh \& Kwon
(2013). We assume that the dust density distribution is continuous from the
inner radius ($R_{in}$) to the outer radius ($R_{out}$) for a spherically
symmetric dust shell. For the dust density distribution, we use the simple
power law equation
\begin{equation}
\rho(r) = \rho_{in} (r/R_{in})^{-2},
\end{equation}
where $\rho_{in}$ is the dust density at $R_{in}$ of the dust shell. The
mass-loss rate is given by
\begin{equation}
\dot{M} = 4 \pi r^2 \rho v_{exp} = 4 \pi R^2_{in} \rho_m v_{exp},
\end{equation}
where $\rho_m$ is the material density at $R_{in}$ which is given by
\begin{equation}
\rho_m \delta = \rho_{in},
\end{equation}
where $\delta$ is the dust-to-gas ratio which is assumed to be 0.01 (e.g., Suh
2014). The mass-loss rate ($\dot{M}$) and dust shell expansion velocity
($v_{exp}$) remain constant.

The inner shell dust temperature ($T_c$) is the dust temperature at $R_{in}$.
The model SEDs are sensitively dependent on $R_{in}$ (or $T_c$). For all AGB
and post-AGB stars, we assume that the dust formation (or condensation)
temperature ($T_c$) is 1000 K (e.g., Suh 1999, 2004).

If dust formation ceases when the post-AGB phase starts, the dust shell begins
to detach (i.e., $R_{in}$ increases for colder $T_c$). For all of the models,
we use the outer radius ($R_{out}$) of the dust shell so that the dust
temperature at $R_{out}$ is lower than 30 K which is comparable to the
temperature of interstellar medium.

We assume that all dust grains are spherical with a uniform radius of 0.1
$\mu$m and the scattering is assumed to be isotropic. We choose 10 $\mu$m as
the fiducial wavelength that sets the scale of the dust optical depth
($\tau_{10}$).

For the central star, we assume that it emits blackbody radiation for a given
temperature and luminosity.

\subsection{AGB stars}

For AGB stars, we use similar assumptions for the dust shell models as those
used for Suh \& Kwon (2011). For O-rich stars, we use the optical constants of
warm and cold silicate grains derived by Suh (1999). We compute models for
eleven dust optical depths ($\tau_{10}$ $=$ 0.005, 0.01, 0.05, 0.1, 0.5, 1, 3,
7, 15, 30, and 40). We use the warm silicate dust grains for low mass-loss rate
O-rich AGB stars (7 models with $\tau_{10}$ $\leq 3$) and the cold silicate
grains for high mass-loss rate O-rich AGB stars (4 models with $\tau_{10}$ $>
3$). For the central star, we assume that the luminosity is $10^4$ $L_{\odot}$
and the stellar blackbody temperature is 2500 K for $\tau_{10}$ $\leq 3$ and
2000 K for $\tau_{10}$ $> 3$.

For C-rich stars, we use the optical constants of amorphous carbon (AMC) grains
derived by Suh (2000). We perform the model calculations for seven dust optical
depths ($\tau_{10}$ $=$ 0.01, 0.1, 1, 2, 3, 5, and 7). For the central star, we
assume that the luminosity is $10^4$ $L_{\odot}$ and the stellar blackbody
temperature is 2300 K for $\tau_{10}$ $\leq 0.1$ and 2000 K for $\tau_{10}$ $>
0.1$.

In Figures 1 - 4, the theoretical model tracks for AGB stars are plotted on IR
2CDs. Suh \& Kwon (2011) presented more various model tracks for AGB stars
using more dust species.

\section{Spectral evolution of Post-AGB stars}

To understand possible evolutionary tracks from AGB stars to PNe on 2CDs, we
investigate the spectral evolution of post-AGB stars using the radiative
transfer model as described in section 3.1. We make simple but reasonable
assumptions on the evolution of properties of the dust shell and the central
star to find theoretical evolutionary tracks of post-AGB stars on 2CDs.

The SEDs of typical post-AGB stars show two distinct components: the cooler
component corresponds to the remnant of the AGB dust shell, and the warmer
component corresponds to the photosphere of the central star (e.g., Volk \&
Kwok 1989). The clear separation of the two flux components would be due to
the detached dust shell because the AGB mass-loss terminated some time ago.
The dynamical time scale (after the cessation of AGB mass-loss) can be
estimated by radiative transfer models. The typical time scales of the
post-AGB phase range from several hundred to a thousand years (e.g., Hrivnak
et al. 1989).

\subsection{The central star}

For the central star, we assume a fixed stellar luminosity and changing
stellar blackbody temperatures during the post-AGB phase. We use relevant
model parameters from the theoretical models of post-AGB stars presented by
van Hoof et al. (1997). Their models are based on the works by Wood (1990),
Vassiliadis \& Wood (1993, 1994), Bl\"ocker \& Sch\"onberner (1991), and
Bl\"ocker (1995a,b). Many assumptions have to be made about the mass-loss
rate, the exact time of the end of the AGB phase, and the outflow velocity.
The duration of the post-AGB phase may be as long as 10$^4$-10$^5$ years for
the less massive objects, but more massive objects may last only a few
decades or centuries in the post-AGB phase (Bl\"ocker 1995b).

In this work, we assume that the AGB-type mass-loss stops (i.e., the AGB phase
ends) when the central star has reached an inferred pulsation period of $P_a$ =
100 days (Bl\"ocker 1995b). When the star has subsequently reached an inferred
pulsation period of $P_b$ = 50 days the post-AGB phase (mass-loss) starts. The
mass-loss rate changes smoothly between the AGB and post-AGB phases.

For $P_a$ = 100 days and $P_b$ = 50 days, the calculated evolutionary time
scales of the post-AGB phase show wide variations (about 38 - 8700 years; see
van Hoof et al. 1997) depending on the core-mass ($M_{core}$) and post-AGB
mass-loss rates. If we assume earlier ends of the AGB phase (a larger value of
$P_a$), the time scale can be significantly increased. Comparison of the
preliminary dust shell model (see section 4.2) results with the observations on
the IR 2CDs indicated that the theoretical post-AGB phase needs to be rather
short (within a thousand years) to match the observations.

We restrict ourselves to two different core-mass models for this paper. For
low-mass O-rich stars and C-rich stars, we use $M_{core}$ = 0.605 $M_{\sun}$
and $L_{*}$ = 6310 $L_{\sun}$ with the post-AGB mass-loss rate at 5 times the
standard value. For high-mass O-rich stars, we use $M_{core}$ = 0.696
$M_{\sun}$ and $L_{*}$ = 11610 $L_{\sun}$ with the post-AGB mass-loss rate at
the standard value. The corresponding model parameters are adopted from Table 1
in van Hoof et al. (1997) for the evolution of the central star in the post-AGB
phase. The time scales of the post-AGB phase are 832 and 174 years for the low
core-mass and high core-mass models, respectively.

For the low core-mass model, Table 3 presents five base models for the dust
shell. For the high core-mass model, Table 4 presents three base models for the
dust shell. We will discuss the dust shell models in section 4.2. For each dust
shell model, the time evolution of the surface temperature ($T_{eff}$) of the
central star at eight points in the post-AGB phase as well as the fixed stellar
luminosity ($L_{*}$) are presented.

\begin{table*}
\centering
\begin{footnotesize}
\caption{Theoretical model parameters of the central star and dust shell for low core-mass post-AGB stars.}
\begin{tabular}{llllllllll}
\hline
\hline
\multicolumn{10}{l}{$M_{core}$ = 0.605 $M_{\sun}$, $L_{*}$ = 6310 $L_{\sun}$, Dust: Silicate, $\dot{M}$=(\textbf{EM}:$3.73\times10^{-9}$; \textbf{ML}:$1.21\times10^{-7}$; \textbf{M}:$1.27\times10^{-5}$) $M_{\odot}/yr$} \\
\hline
\textbf{Base Model} & $T_{eff}$ (K) & $t_{pa}$ (years) & $R_{in}$ (AU) & $\rho_{in}$ (g/cm$^{3}$) & $\tau_{10}$ & $T_c$ (K) & \textbf{Derived Model(s)}   \\
\hline
\textbf{EM} &	6042	&	0	&	1.70E+01	&	1.93E-21	&	1.00E-03	&	1000	& \textbf{EM2(75)}: $v_{exp}$=75 \\
$v_{exp}$=15	&	6500	&	113	&	3.75E+02	&	3.98E-24	&	4.59E-05	&	205	& \textbf{EM2(150)}: $v_{exp}$=150 	\\
$\tau_{i} $=0.001 &	7000	&	178	&	5.80E+02	&	1.66E-24	&	2.96E-05	&	174	& $\tau_{i} $=0.001    	\\
	&	10000	&	241	&	7.80E+02	&	9.18E-25	&	2.20E-05	&	168	&   		\\
	&	15000	&	325	&	1.05E+03	&	5.10E-25	&	1.64E-05	&	164	&   		\\
	&	20000	&	452	&	1.45E+03	&	2.66E-25	&	1.18E-05	&	152	&  		\\
	&	25000	&	625	&	1.99E+03	&	1.40E-25	&	8.53E-06	&	138	& 			\\
	&	30000	&	832	&	2.65E+03	&	7.95E-26	&	6.44E-06	&	126	& 			\\
\hline
\textbf{ML}  &	6042	&	0	&	1.90E+01	&	5.00E-20	&	2.90E-02	&	1000 &	\textbf{ML2(75)}: $v_{exp}$=75	 		\\
$v_{exp}$=15	&	6500	&	113	&	3.77E+02	&	1.27E-22	&	1.47E-03	&	205 & \textbf{ML2(150)}: $v_{exp}$=150   	\\
$\tau_{i}$=0.029	&	7000	&	178	&	5.82E+02	&	5.32E-23	&	9.51E-04	&	174 & \textbf{ML2(300)}: $v_{exp}$=300 	\\
 &	10000	&	241	&	7.82E+02	&	2.95E-23	&	7.09E-04	&	168     &	$\tau_{i}$=0.029 	\\
	&	15000	&	325	&	1.05E+03	&	1.65E-23	&	5.32E-04	&	164 &			  	\\
	&	20000	&	452	&	1.45E+03	&	8.59E-24	&	3.82E-04	&	152 &			  	\\
	&	25000	&	625	&	2.00E+03	&	4.53E-24	&	2.78E-04	&	138 &			 	\\
	&	30000	&	832	&	2.65E+03	&	2.57E-24	&	2.08E-04	&	126 &			 	\\
\hline	
\textbf{M}	&	6042	&	0	&	2.47E+01	&	3.10E-18	&	2.35E+00	&	1000 & \textbf{M2(30)}: $v_{exp}$=30 \\
$v_{exp}$=15   &	6500	&	113	&	3.82E+02	&	1.29E-20	&	1.51E-01	&	241 & $\tau_{i}$=2.35 \\
$\tau_{i}$=2.35	&	7000	&	178	&	5.89E+02	&	5.46E-21	&	9.88E-02	&	198 & 	 	\\
	&	10000	&	241	&	7.88E+02	&	3.04E-21	&	7.36E-02	&	190 &			 		\\
	&	15000	&	325	&	1.05E+03	&	1.70E-21	&	5.48E-02	&	180 &			 		\\
	&	20000	&	452	&	1.46E+03	&	8.91E-22	&	3.99E-02	&	162 &			 		\\
	&	25000	&	625	&	2.00E+03	&	4.70E-22	&	2.88E-02	&	144 &			 		\\
	&	30000	&	832	&	2.66E+03	&	2.67E-22	&	2.17E-02	&	130 &			 		\\
\hline
\hline
\multicolumn{10}{l}{$M_{core}$ = 0.605 $M_{\sun}$, $L_{*}$ = 6310 $L_{\sun}$, Dust: Amorphous Carbon, $\dot{M}$=(\textbf{CL}:$7.85\times10^{-8}$; \textbf{C}:$1.68\times10^{-5}$) $M_{\odot}/yr$} \\
\hline
\textbf{Base Model} & $T_{eff}$ (K) & $t_{pa}$ (years) & $R_{in}$ (AU) & $\rho_{in}$ (g/cm$^{3}$) & $\tau_{10}$ & $T_c$ (K) & \textbf{Derived Model(s)}   \\
\hline
\textbf{CL}	&	6042	&	0	&	1.78E+01	&	3.70E-20	&	1.00E-02	&	1000 &	\textbf{CL2(30)}: $v_{exp}$=30	\\
$v_{exp}$=15  &	6500	&	113	&	3.75E+02	&	8.32E-23	&	4.77E-04	&	298 & \textbf{CL2(150)}: $v_{exp}$=150		\\
$\tau_{i}$=0.01 &	7000	&	178	&	5.81E+02	&	3.47E-23	&	3.08E-04	&	255 & \textbf{CL2(300)}: $v_{exp}$=300		\\
  &	10000	&	241	&	7.80E+02	&	1.93E-23	&	2.30E-04	&	241 &	$\tau_{i}$=0.01	 	\\
  &	15000	&	325	&	1.05E+03	&	1.07E-23	&	1.72E-04	&	222 &				 	\\
	&	20000	&	452	&	1.45E+03	&	5.59E-24	&	1.24E-04	&	197 &			 	\\
	&	25000	&	625	&	2.00E+03	&	2.94E-24	&	8.99E-05	&	175 &			 	\\
	&	30000	&	832	&	2.65E+03	&	1.67E-24	&	6.77E-05	&	156 &			  	\\
\hline
\textbf{C}  &	6042	&	0	&	1.83E+01	&	7.50E-18	&	2.11E+00	&	1000   & \textbf{C2(30)}: $v_{exp}$=30		\\
$v_{exp}$=15  &	6500	&	113	&	3.76E+02	&	1.78E-20	&	1.03E-01	&	311    & $\tau_{i}$=2.11  	\\
$\tau_{i}$=2.11 &	7000	&	178	&	5.82E+02	&	7.43E-21	&	6.67E-02	&	266 &     	\\
 &	10000	&	241	&	7.81E+02	&	4.12E-21	&	4.96E-02	&	251 &			 		\\
 &	15000	&	325	&	1.05E+03	&	2.29E-21	&	3.71E-02	&	232 &			 		\\
	&	20000	&	452	&	1.45E+03	&	1.20E-21	&	2.68E-02	&	207 &			 	\\
	&	25000	&	625	&	2.00E+03	&	6.31E-22	&	1.94E-02	&	182 &			 	\\
	&	30000	&	832	&	2.65E+03	&	3.57E-22	&	1.45E-02	&	162 &			 	\\
\hline
\end{tabular}
\begin{flushleft}
Dust shell parameters are presented for base models (see section 4.2). The unit for $v_{exp}$ is km/sec. Note that the model tracks for
\textbf{EM2(75)}, \textbf{EM2(150)}, \textbf{ML2(150)},\textbf{ML2(150)}, \textbf{ML2(300)}, and \textbf{CL2(300)}
are plotted only on the $IRAS$ 2CD (Figure 1).
\end{flushleft}
\end{footnotesize}
\end{table*}

\begin{table*}
\centering
\begin{footnotesize}
\caption{Theoretical model parameters of the central star and dust shell for high core-mass post-AGB stars.}
\begin{tabular}{llllllllll}
\hline
\hline
\multicolumn{10}{l}{$M_{core}$ = 0.696 $M_{\sun}$, $L_{*}$ = 11610 $L_{\sun}$, Dust: Silicate, $\dot{M}$=(\textbf{OL}:$8.64\times10^{-5}$; \textbf{O}:$3.40\times10^{-4}$; \textbf{T}:$8.38\times10^{-4}$) $M_{\odot}/yr$} \\
\hline
\textbf{Base Model} & $T_{eff}$ (K) & $t_{pa}$ (years) & $R_{in}$ (AU) & $\rho_{in}$ (g/cm$^{3}$) & $\tau_{10}$ & $T_c$ (K) & \textbf{Derived Model}    \\
\hline
\textbf{OL}	&	6846	&	0	&	2.54E+01	&	2.00E-17	&	1.55E+01	&	1000 	& \textbf{OL2(75)}: $v_{exp}$=75 \\
$v_{exp}$=15  &	7000	&	29	&	1.17E+02	&	9.40E-19	&	3.38E+00	&	558 	& $\tau_{i}$=15.5			 	\\
$\tau_{i}$=15.5 &	7500	&	62	&	2.22E+02	&	2.63E-19	&	1.79E+00	&	415 &	    	 	\\
    &	10000	&	84	&	2.91E+02	&	1.52E-19	&	1.36E+00	&	392 	&			 	\\
	&	15000	&	111	&	3.77E+02	&	9.10E-20	&	1.05E+00	&	369 	&			 	\\
	&	20000	&	129	&	4.34E+02	&	6.86E-20	&	9.13E-01	&	354 	&			 		\\
	&	25000	&	151	&	5.03E+02	&	5.10E-20	&	7.86E-01	&	334 	&			 		\\
	&	30000	&	174	&	5.76E+02	&	3.89E-20	&	6.86E-01	&	314 	&			 		\\
\hline
\textbf{O}    &	6846	&	0	&	3.00E+01	&	5.65E-17	&	5.20E+01	&	1000	& \textbf{O2(75)}: $v_{exp}$=75 \\
$v_{exp}$=15 &	7000	&	29	&	1.22E+02	&	3.43E-18	&	1.21E+01	&	501	    & $\tau_{i}$=52  \\
$\tau_{i}$=52  &	7500	&	62	&	2.26E+02	&	9.94E-19	&	6.48E+00	&	373	&		\\
    &	10000	&	84	&	2.96E+02	&	5.81E-19	&	4.96E+00	&	343	    &			\\
    &	15000	&	111	&	3.81E+02	&	3.50E-19	&	3.84E+00	&	321	&				\\
	&	20000	&	129	&	4.38E+02	&	2.65E-19	&	3.34E+00	&	310	&			\\
	&	25000	&	151	&	5.08E+02	&	1.97E-19	&	2.89E+00	&	296	&			\\
	&	30000	&	174	&	5.81E+02	&	1.51E-19	&	2.53E+00	&	283	&			\\
\hline
\textbf{T}  &	6846	&	0	&	3.75E+01	&	8.90E-17	&	1.02E+02	&	1000 	& \textbf{T2(75)}: $v_{exp}$=75 	\\
$v_{exp}$=15	&	7000	&	29	&	1.29E+02	&	7.49E-18	&	2.97E+01	&	485 	&  $\tau_{i}$=102	\\
$\tau_{i}$=102  &	7500	&	62	&	2.34E+02	&	2.29E-18	&	1.65E+01	&	360 	&	    	 \\
	&	10000	&	84	&	3.03E+02	&	1.36E-18	&	1.26E+01	&	320 	&			  	\\
    &	15000	&	111	&	3.89E+02	&	8.28E-19	&	9.87E+00	& 290   &		 \\
	&	20000	&	129	&	4.46E+02	&	6.30E-19	&	8.60E+00	&	276 	&			\\
	&	25000	&	151	&	5.15E+02	&	4.71E-19	&	7.42E+00	&	263 	&			\\
	&	30000	&	174	&	5.88E+02	&	3.62E-19	&	6.50E+00	&	252 	&			\\
\hline
\end{tabular}
\begin{flushleft}
Dust shell parameters are presented for base models (see section 4.2). The unit for $v_{exp}$ is km/sec.
\end{flushleft}
\end{footnotesize}
\end{table*}

\subsection{The dust shell}

For the dust shell, we use dust radiative transfer model described in section
3.1. We use the same dust opacity functions as those used for AGB stars. we
use the optical constants of cold silicate dust grains derived by Suh (1999)
for O-rich post-AGB stars. For C-rich post-AGB stars, we use the optical
constants of amorphous carbon (AMC) dust grains derived by Suh (2000).

For C-rich post-AGB stars, we use AMC rather than graphite which is widely
used for interstellar dust or young stellar objects. It is believed that AMC
is the main dust component for C-rich post-AGB stars (e.g., Hony et al. 2003;
Cerrigone et al. 2009) as well as for C-rich AGB stars. Unlike AGB stars,
C-rich post-AGB stars show dust features from hydrogenated amorphous carbon
and PAH as well as SiC and MgS (e.g., Justtanont et al. 1996a).

The typical dust shell expansion velocity ($v_{exp}$) is about 15 km/sec for
AGB stars (e.g., Loup et al. 1993; Suh 2014). Even though it is not clear
whether $v_{exp}$ can be increased during the AGB phase or post-AGB phase,
$v_{exp}$ could be higher in the post-AGB phase. Slijkhuis et al. (1991)
measured the outflow velocities up to $v_{exp}$ = 400 km/sec for a post-AGB
star ($IRAS$ 08005-2356). We use various possible values of $v_{exp}$ (15,
30, 75, 150, and 300 km/sec) for post-AGB stars.

Uncertainty on the mass-loss rate remains. A thermal pulse in the AGB phase
can change the mass-loss rate abruptly (e.g., Renzini 1981; Wood 1990;
Vassiliadis \& Wood 1993, 1994) from the continuous analytical laws which
were used for the central star (see section 4.1). The effect of the thermal
pulse on $T_{eff}$ would be minor, but the effect on $L_{*}$ could be
significant. However, a change in $L_{*}$ does not affect the shape of the
output spectra (or color), it only affects the overall energy output.
Therefore, we may ignore the effect of the thermal pulse on the mean
properties ($L_{*}$, $T_{eff}$) of the central stars for a large sample of
evolving stars.

However, a change in the mass-loss rate of the dust shell due to the thermal
pulse affects the observed SED significantly. The various effects of the
thermal pulse on a dust shell (a superwind and/or chemical transition from O to
C) in the AGB phase are well investigated (e.g., Groenewegen et al. 1995;
Justtanont et al. 1996b; Suh \& Jones 1997). Considering the effects of the
thermal pulses, we use various mass-loss rates of the dust shell (rather than a
fixed one) at the end of the AGB phase.

To make a simplified dust shell model as described in section 3.1, we assume
that the mass-loss rate and $v_{exp}$ of the dust shell at the end of the AGB
phase remain constant during the period between the end of AGB phase and the
start of the post-AGB phase. And we assume that AGB dust formation (at $T_c$
= 1000 K) continues during the period between the end of AGB phase and the
start of the post-AGB phase because dust formation properties during the
period can not be easily clarified.

We assume that the AGB dust formation (at $T_c$ = 1000 K) ceases with the
start of the post-AGB phase. $t_{pa}$ is defined as the time scale since the
start of the post-AGB phase. With the start of post-AGB phase, the inner
radius ($R_{in}$) of the dust shell increases with time ($t_{pa}$):
\begin{equation}
R_{in}=R_{0}+v_{exp}t_{pa},
\end{equation}
where $R_{0}$ is $R_{in}$ at $t_{pa}$=0. As the post-AGB star evolves, $T_{c}$
gets colder as the dust shell detaches. For the dust shell, we assume various
initial dust optical depths at the start of the post-AGB phase so that the
corresponding mass-loss rates (Equation 3) are reasonably in the range of
observed values of AGB stars.

We make various dust shell models with different mass-loss rates ($\dot{M}$)
for each core-mass model of the central star. For the low core-mass model, we
present three base models (EM, ML, M) for low-mass O-rich post-AGB stars and
two base models (CL and C) for C-rich post-AGB stars in Table 3. For the high
core-mass model, we present three base models (OL, O, and T) for high-mass
O-rich post-AGB stars in Table 4. For all dust shell models, the expansion
velocity ($v_{exp}$) and mass-loss rate ($\dot{M}$) of the dust shell remain
constant during the whole post-AGB phase (see section 3.1).

In Tables 3 and 4, the mass-loss rates are presented for base models. We assume
that $v_{exp}$ = 15 km/sec for all base models. For each base model, dust shell
parameters ($R_{in}$, $\rho_{in}$, $\tau_{10}$, and $T_c$) at eight points of
time evolution in the post-AGB phase are listed. $\tau_{i}$ is defined as the
initial dust optical depth ($\tau_{10}$) at the start of the post-AGB phase. As
the post-AGB phase evolves, $\tau_{10}$ and the output parameter $T_c$ (the
inner shell dust temperature) decrease while $R_{in}$ increases with time
($t_{pa}$).

For each base model, we make model calculations for various derived models with
faster dust shell expansion velocities ($v_{exp}$ = 30, 75, 150, and 300
km/sec). From the base model EM, we have two derived models EM2(75) and
EM2(150) using $v_{exp}$ = 75 and 150 km/sec, respectively. The derived model
uses the same model parameters (e.g., $R_{0}$ and $\tau_{i}$) and produces the
same model SED as the base model at $t_{pa}$=0. At later post-AGB phases, the
derived model uses the same stellar parameters as the base model but different
dust shell parameters ($R_{in}$, $\rho_{in}$, and $\tau_{10}$) because of
higher $v_{exp}$ (see Equation 5). Note that $\dot{M}$ for the derived model
($\dot{M}_d$) is given by
\begin{equation}
\dot{M}_d=\dot{M}_{b} \frac {v_{exp}}{15},
\end{equation}
where $\dot{M}_b$ is $\dot{M}$ for the base model (see Equation 3).

\subsection{Model SEDs}

Figure 5 shows the model SEDs for post-AGB stars for six different models. At
the beginning of the post-AGB phase, the SEDs look like those of AGB stars. As
the star evolves to a later stage, the SED shows more distinct double
components, which are typical for post-AGB stars.

Observed SEDs of many post-AGB stars show prominent silicate features (e.g.,
Hrivnak et al. 1989; Cerrigone et al. 2009) as well as the model SEDs for
O-rich stars. The silicate features become very weak at the end of the
post-AGB phase especially for the models with high $v_{exp}$. We will compare
the theoretical model tracks with observations on IR 2CDs in section 5.

\begin{figure*}
\plotsix{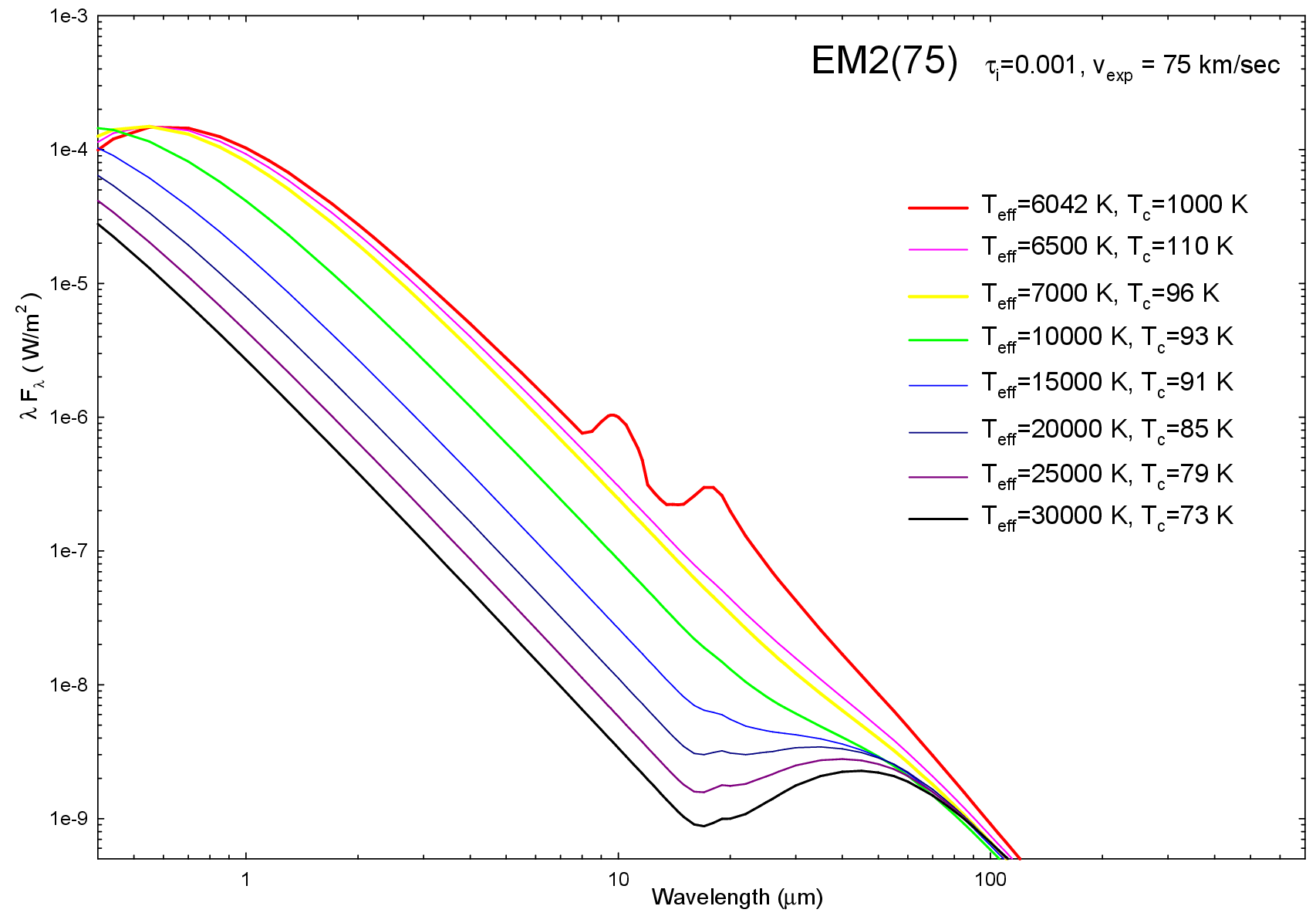}{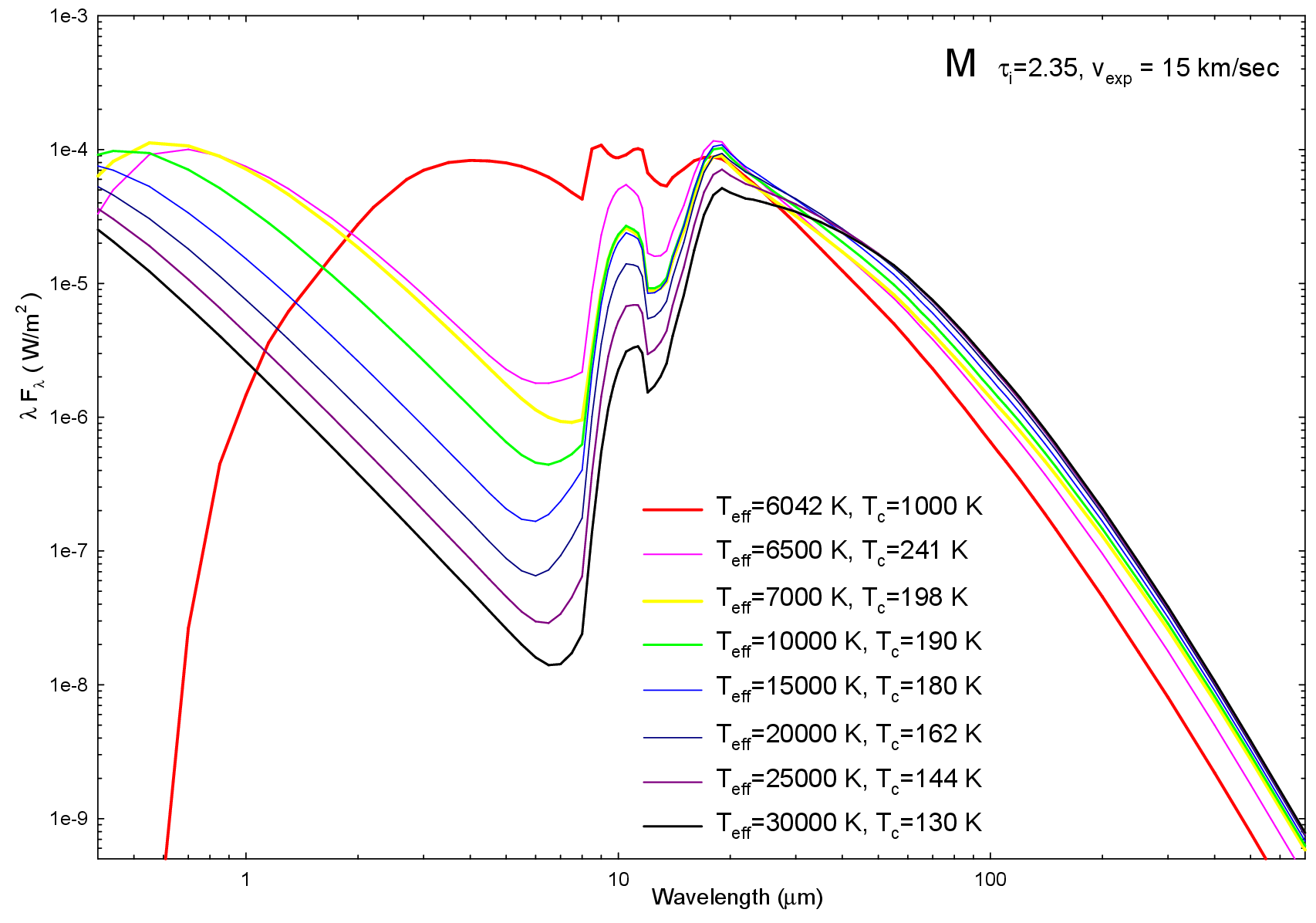}{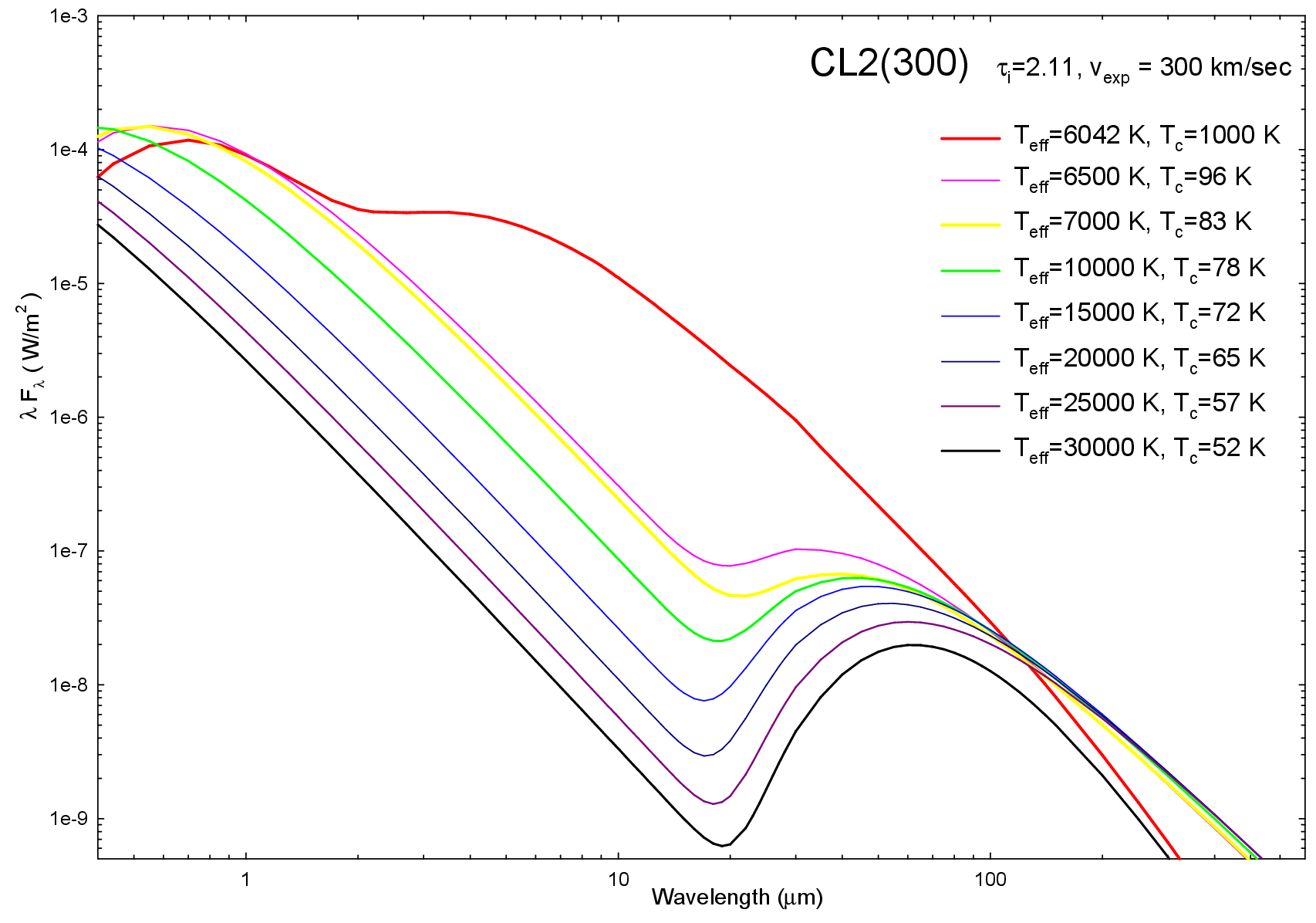}{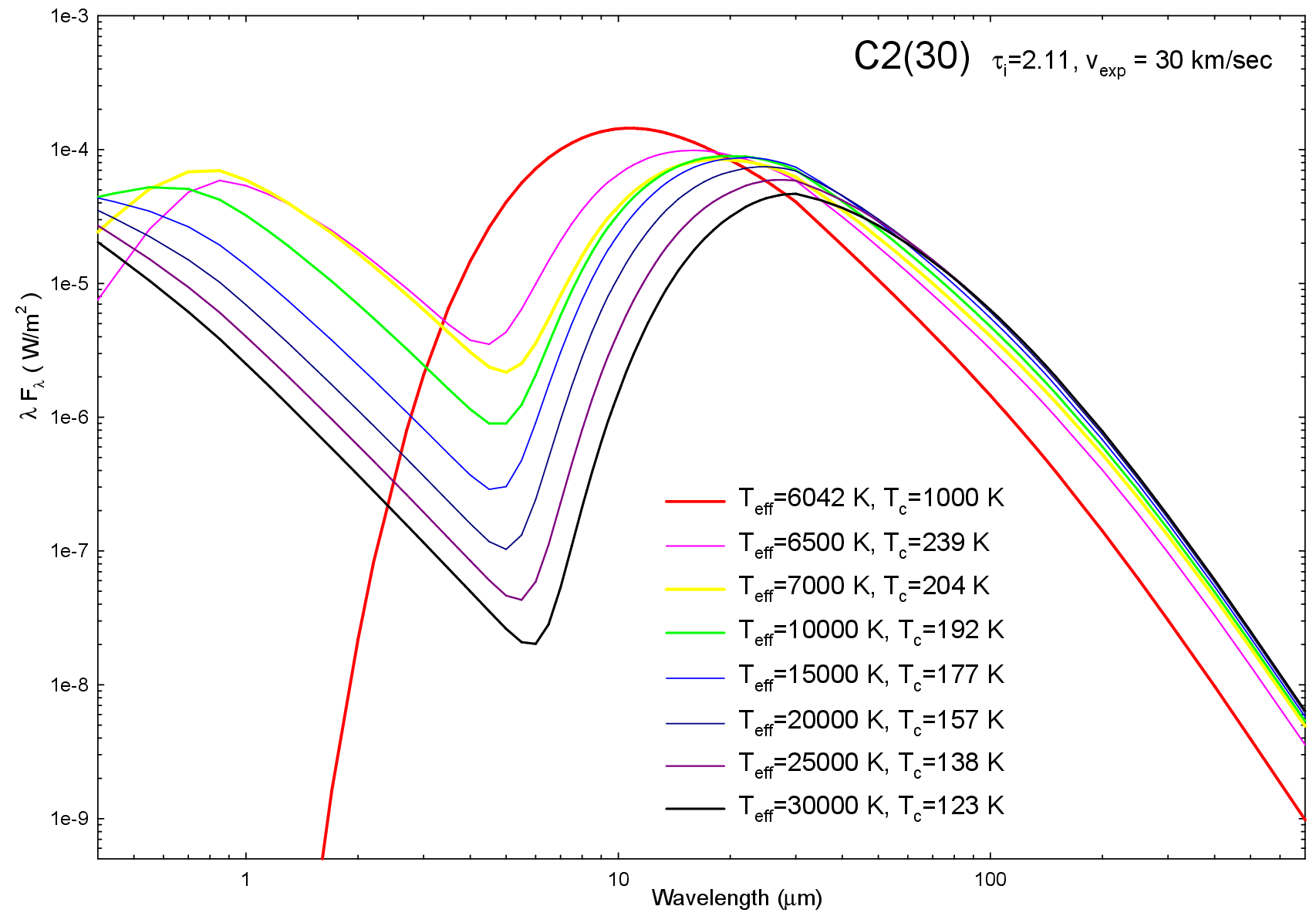}{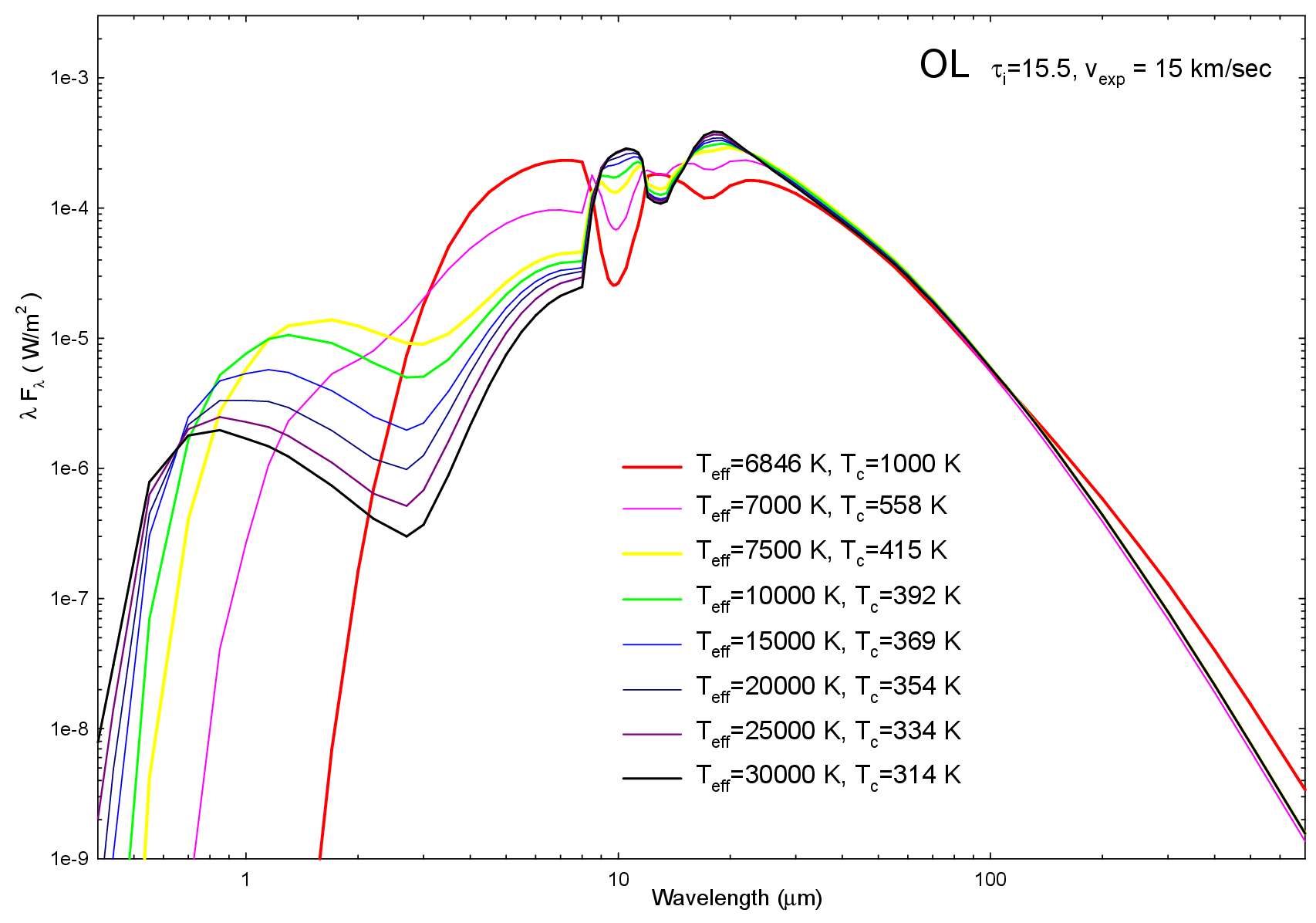}{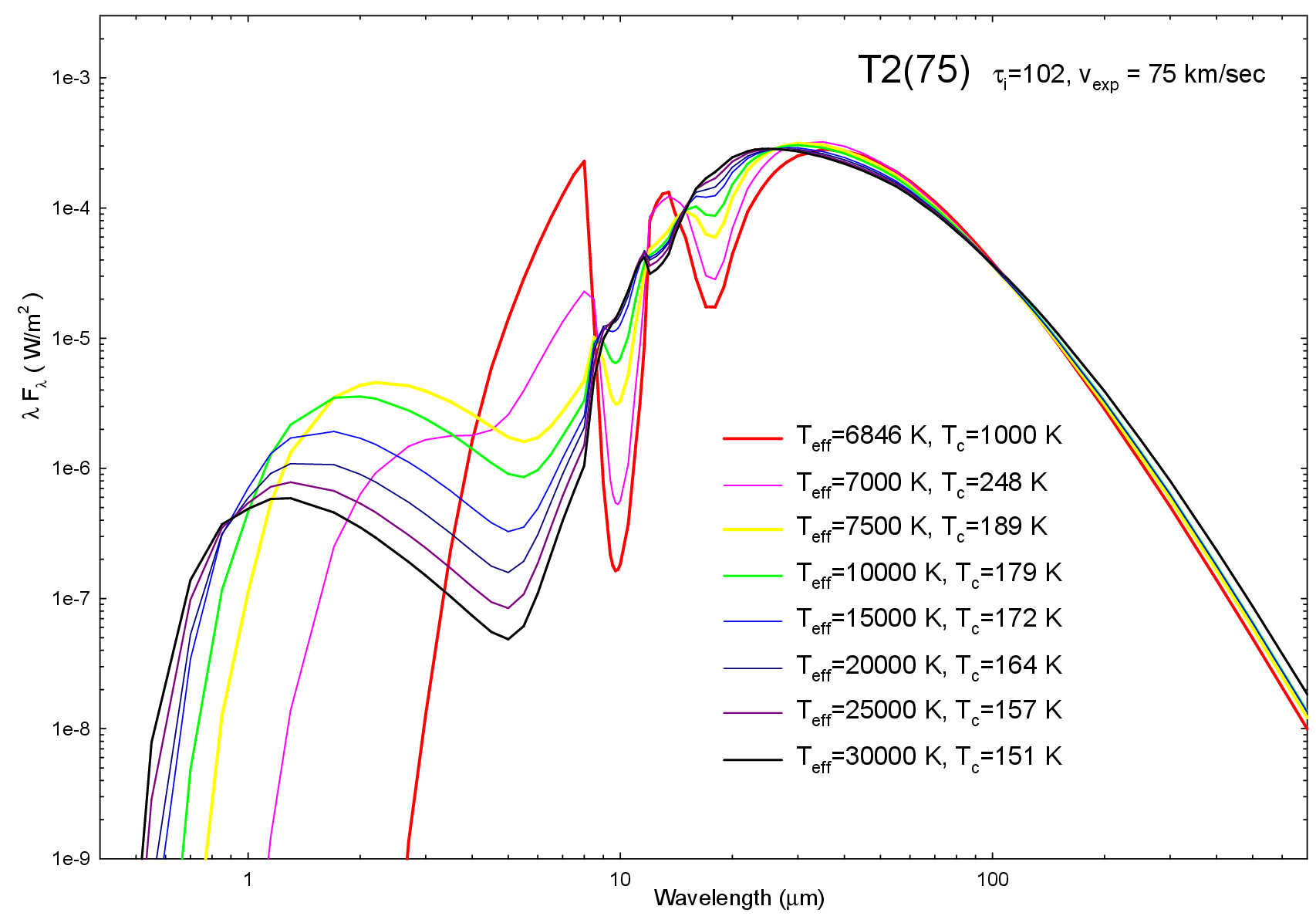} \caption{Model SEDs for post-AGB stars.
We assume that the distance is 1 pc. See Tables 3 and 4 for detailed model parameters.}
\end{figure*}

\section{Comparison on IR 2CDs}

Figures 1 - 4 show $IRAS$, NIR, $AKARI$, and $MSX$ 2CDs for AGB stars, post-AGB
stars, and PNe compared with theoretical models. The small symbols are the
observational data (see section 2.4) and the lines with large symbols are
theoretical model tracks for AGB stars (section 3.1). The lines with filled
symbols are theoretical model tracks for post-AGB stars (section 4).

For AGB stars, theoretical models for a range in dust shell optical depth is
presented (see section 3.1). The objects in the upper-right regions on any
2CDs have thick dust shells with large optical depths for AGB stars. The
locations of S stars are restricted to the regions of thin dust shells
(lower-left regions) on any IR 2CDs. For AGB stars, see Suh \& Kwon (2011)
for a detailed discussion on the comparison of the theoretical models with
the observations on various IR 2CDs.

To find possible evolutionary tracks from AGB stars to PNe on 2CDs, we present
the theoretical model tracks for the evolving post-AGB phase (see Tables 3 and
4 for detailed model parameters). For each post-AGB model track, the point at
the start of the post-AGB phase is indicated by the tip of an arrow and the
base model name. The base or derived model name is labeled at the end point of
the model track.

We find that the theoretical dust shell model tracks for post-AGB stars
roughly coincide with the densely populated observed points of post-AGB stars
and PNe on various IR 2CDs. Even though some discrepancies are inevitable,
the end points of the post-AGB model tracks are generally converged to the
region of the observed points of PNe on most 2CDs.

If our assumptions on other parameters are right, the time scale of the
post-AGB phase (832 years) for low core-mass models (EM, ML, M, CL, and C)
looks reasonable. The models for high core-mass stars (models OL, O, and T),
for which the time scale is even shorter (174 years), match the observations on
IR 2CDs fairly well with various expansion velocities.

Unlike AGB stars, C-rich post-AGB stars or PNe typically show PAH dust
emission features at 3.3, 6.2, 7.7, 8.6, 11.3, and 12.5 $\mu$m (e.g.,
Justtanont et al. 1996a; Cerrigone et al. 2009). This could be due to UV
radiation from the hot central stars that have evolved past the AGB. The
observed colors [8.28]$-$[14.65] and [12]$-$[25] could be bluer than the
theoretical model colors which do not consider PAH dust.

The effects of very thin dust shell models (EM, ML, and CL) or very thick
models (T) can be more clearly presented on the $IRAS$ 2CD than any other 2CDs
considered in this work. This is because the [25]$-$[60] color, which uses
longer wavelengths, can represent the evolving SED better than any other colors
(see Figure 5). Other model colors can be more easily saturated to certain
values because they can be more severely affected by the boundary between two
components (the central star and the dust shell) and/or silicate dust features.
Because $K_{s}$$-$[12], [9]$-$[12], and [8.28]$-$[14.65] model colors are
rapidly converged to certain values for very small dust optical depths, the IR
2CDs in Figures 2-4 are not useful to investigate the thin dust shell models
compared with the $IRAS$ 2CD (Figure 1). Therefore, the model tracks for
EM2(75), M2(150), ML2(150), ML2(150), ML2(300), and CL2(300) are plotted only
on the $IRAS$ 2CD (Figure 1; see Table 3).

\subsection{The $IRAS$ 2CD}

Figure 1 shows the $IRAS$ 2CDs using [25]$-$[60] versus [12]$-$[25] for AGB
stars and post-AGB stars compared with theoretical models. The eight regions
and the curved green line which is the evolutionary track for late-type stars
were defined by van der Veen \& Habing (1988). The theoretical model tracks for
AGB stars and the start points of the post-AGB model tracks roughly coincide
with the evolutionary track for late-type stars which were defined by van der
Veen \& Habing (1988).

Brown dashed regions on the $IRAS$ 2CD indicate the two groups of post-AGB
stars, which are designated as LI (Left of $IRAS$, blue group) and RI (Right of
$IRAS$, red group) (Sevenster 2002). Most post-AGB objects are distinguished by
very red MIR colors (RI region), but there is also a group of very blue sources
with a strong 60 $\mu$m excess (LI region). A large number of post-AGB and PNe
samples are mainly distributed in the RI region compared with limited samples
of the LI region (e.g., Yoon et al. 2014). Sevenster (2002) indicated that LI
objects are more massive and have higher outflow velocities than RI objects. LI
objects may go through an irregular mass-loss phase, preceding the bipolar PN
stage. On the other hand, RI objects are the traditional post-AGB sources (van
der Veen \& Habing 1988), turned right from the evolutionary sequence into
$IRAS$ region V and beyond.

The RI objects can be explained by a normal evolutionary sequence for which
stars undergo dust shell detachments at the normal AGB termination. For all of
the post-AGB model tracks with moderate $v_{exp}$ (15 - 75 km/sec), the end
points are located in this RI region. This is in accord with the fact that most
of the observed PNe are converged to the RI region (see Figure 1). This trend
is more noticeable when we consider only the good quality observational data of
PNe which are plotted by thicker symbols on the 2CD.

The LI objects could be explained by with extremely thin dust shells (models
EM2, ML2, and CL2 with faster $v_{exp}$; see Figure 1). However, the
theoretical model used for this work may not be applicable because most of
the LI objects are believed to be non-spherical systems (precursors of
high-mass, bipolar objects or binary systems).

If we assume that related stars are single and nearly spherical objects, the
LI objects as well as the objects in upper-left region (VI) are likely to be
hot objects with very thin dust shells. We find that the thin dust shell
models (EM(2), ML(2), and CL(2)) with various $v_{exp}$ at the beginning of
the post-AGB phase can explain the post-AGB stars in the lower-left region
(I, II, III, and VII) on the $IRAS$ 2CD. For the post-AGB stars and PNe in
the upper-left region (VI and LI), the model requires a very thin dust shell
(models EM2, ML2, and CL2) with higher $v_{exp}$ (75-300 Km/sec) and a hotter
central star which is more evolved in the post-AGB phase. This requirement of
a thin dust shell model may look to be in contradiction to the idea that a
star at the end of the AGB phase would have a very thick dust shell. This
contradiction could be explained by the effect of the chemical transition
from O to C in the AGB phase (see section 6.3).

\subsection{NIR, $AKARI$, and $MSX$ 2CDs}

Figure 2 shows $IRAS$-2MASS 2CDs using [12]$-$[25] versus $K_{s}$$-$[12].
Compared with the $IRAS$ 2CD, the boundaries that separate O-rich, C-rich and S
stars are clearer for AGB stars. The locations of observed PNe are converged
toward upper-middle region. Unlike other 2CDs, the locations of the end points
for the post-AGB model tracks look to show systematic deviation from the
densely populated region of the observed PNe. $K_{s}$$-$[12] model colors are
generally redder than the observed colors of PNe.

Post-AGB stars and PNe typically show H$_{2}$ line emission in the $K_{s}$ band
(e.g., Davis et al. 2003; Froebrich et al. 2011). The 2MASS data in the $K_{s}$
band (2.17 $\mu$m) may show this effect. If the effect of H$_{2}$ emission on
the $K_{s}$ band is stronger than the one from the PAH emission on the [12]
band, the observed $K_{s}$$-$[12] colors could be bluer compared with the
theoretical model colors which do not consider the H$_{2}$ line emission.
Another possible reason for the difference could be that the 2MASS flux may
include only the central star while the $IRAS$ fluxes may include the extended
post-AGB dust envelope (see section 2).

Figure 3 shows $AKARI$-$IRAS$ 2CDs using [9]$-$[12] versus [12]$-$[25].
Generally, the end points of the O-rich and C-rich post-AGB model tracks are
converged to the region of observed PNe. Compared with other 2CDs, it is
difficult find any boundaries that separate O-rich AGB stars and C-rich AGB
stars. But the boundary that separates AGB stars and PNe is clear with the aid
of the distinguishing [12]$-$[25] color. The model [9]$-$[12] colors can be
affected by some dust species (alumina, SiC, and PAH) which are not considered
in this work.

Figure 4 plots $MSX$ 2CDs using [8.28]$-$[14.65] versus [14.65]$-$[21.34].
Unlike other 2CDs, the base models (OL, O, and T) for high core-mass stars can
match the observations better than the models with faster $v_{exp}$. Generally,
this $MSX$ 2CD looks to require a faster evolution of the post-AGB phase than
other 2CDs. The [14.65]$-$[21.34] color is useful to distinguish different
classes of objects on the 2CD. For the [8.28]$-$[14.65] color, the PAH emission
feature at 8.6 $\mu$m from post-AGB stars and PNe could make the observed
colors bluer (downward on the 2CD) compared with the theoretical models.

\section{Discussion}

\subsection{Uncertainties and alternative models}

To investigate spectral evolution from AGB stars to PNe, we have performed
radiative transfer model calculations for the evolving dust shells and
central stars in the post-AGB phase. We have assumed that the central star
emits blackbody radiation. Uncertainties remain in a number of aspects.

The spherically symmetric dust shell models used for this work could be too
simple for post-AGB stars because their shapes are often believed to be
non-spherical (e.g., bipolar or disk-like objects). However, the theoretical
models for non-spherical dust envelopes (disks or toruses) which are applicable
to the highly non-spherical objects should have a much larger number of assumed
model parameters which can be difficult to be compared with a large sample of
observed objects.

Our assumption of the blackbody radiation for the central star inside a dust
shell could be too simple especially for post-AGB stars for which gas-phase
radiation from the outer shell can be important. For the evolution of a
post-AGB star with an expanding gas shell, Volk (1992) used the output of the
photoionization code CLOUDY (Ferland 1993) as input for a dust shell model
(CSDUST3). But this could have a problem of inconsistent treatment of dust
temperatures. van Hoof et al. (1997) used the CLOUDY code considering dust
emission more consistently for an optically thin dust shell. However, most
dust shell models at the beginning of the post-AGB phase require large dust
optical depths.

For dust opacity, we have not considered some dust species: alumina, SiC, and
PAH. Suh \& Kwon (2011) used a mixture of alumina and silicate for O-rich AGB
stars and a mixture of SiC and AMC in modelling C-rich AGB stars for various
model tracks. We also used the same mixtures for testing post-AGB models,
which produced different model tracks but the overall match to the
observations looked similar. Unlike AGB stars, C-rich post-AGB stars or PNe
typically show PAH dust features. We have discussed possible effects of PAH
on IR 2CDs in section 5.

\subsection{Comparison with other works on post-AGB stars}

Even though both our dust shell model and van Hoof et al. (1997)'s work
assume a spherically symmetric continuous dust shell, van Hoof et al. (1997)
used significantly different schemes in their radiative transfer code
(CLOUDY) from those used for this work (RADMC-3D). The strong point is that
their model can consider some gas-phase radiation (bound-free emission for
hot models). As commented in the paper, the weak point is that the CLOUDY
model for a thick dust shell could be unreliable. van Hoof et al. (1997)
provided many meaningful results, but some of their results need to be
considered more carefully because dust optical depths at the beginning of the
post-AGB phase are generally large.

On the $IRAS$ 2CD, van Hoof et al. (1997) presented the red loop model track
which causes the [12]-[25] color to loop back toward the blue, before the star
resumes its normal evolution to the red. Some of our models (O and T) also show
similar tracks, but they do not cover deep into the LI region. Sevenster (2002)
tried to explain the LI objects on the $IRAS$ 2CD (see section 5.1) using the
leftward model track in the early phases of the red loop track. However, their
dust shell model (CLOUDY) could be unreliable because the corresponding dust
optical depths (for the leftward model track) are very large. Volk (1992) and
Ortiz et al. (2005), who used the output of the photoionization code (CLOUDY)
as input for a dust shell model (CSDUST3), also did not find the red loop model
track.

If our assumptions are right, the LI region of the $IRAS$ 2CD can be covered
by the model tracks with thin dust shells at the starts of their post-AGB
phases and larger detachments which can be achieved by higher expansion
velocities (e.g., models EM2, ML2, and CL2 in this work). These models make
upward model tracks rather than leftward (see Figure 1). Of course, any
spherically symmetric dust shell models would be inappropriate for highly
non-spherical objects in the LI region.

For the low core-mass model, we also used the model for $M_{core}$ = 0.605
$M_{\sun}$ with the post-AGB mass loss rate at the standard value (see section
4.1). The corresponding dynamical time scale of the post-AGB phase is 2093
years. This model produced the model results which are too far off from the
observed points of post-AGB stars and PNe on all 2CDs. Volk (1992) and Ortiz et
al. (2005) investigated spectral evolution of post-AGB stars using the similar
scheme for the evolution of the central star. They also showed that a fast
evolution (the time scale of less than about two thousand years) was required
to match the observations on 2CDs. Because they assumed an earlier start of the
post-AGB phase, the dynamical time scales from Ortiz et al. (2005) are
generally larger than those used for this work.

\subsection{The chemical transition in the AGB phase}

As stars evolve into the thermal pulsing AGB phase, the abundances of some
elements in the stellar atmosphere may change by the episodic third dredge-up
process after each thermal pulse. When AGB stars of intermediate mass range
go through carbon 'dredge-up' processes, and thus the abundance of carbon is
larger than that of oxygen, O-rich dust grain formation ceases and the stars
become visual carbon stars. After that phase, carbon-rich dust grains start
forming and the stars evolve into infrared carbon stars with thick C-rich
dust envelopes and very high mass-loss rates (e.g., Iben 1981; Chan \& Kwok
1990; Suh 2000).

S stars are generally regarded as intermediate between M-type and carbon stars
in their properties. However, this M-S-C evolutionary sequence can be different
depending on the mass and metallicity; some stars may remain in S-type until a
next thermal pulse but some stars may skip the S-type star phase (e.g.,
Groenewegen et al. 1995). SC or CS stars are likely to be in a transitional
phase from an S star to a carbon star. Generally, dust formation in envelopes
of S stars is less efficient because of the lack of free O or C to form dust.
Most S stars show very weak dust (silicate or carbon) emission features and low
mass-loss rates (e.g., Hony et al. 2009; Smolders et al. 2012) and their
locations are restricted to the regions of thin dust shells (lower-left
regions) on any IR 2CDs (see Figures 1 - 4).

To reproduce the post-AGB stars in lower-left and upper-left regions on the
$IRAS$ 2CD, we need to use very thin dust shell models with higher $v_{exp}$ at
the beginning of the post-AGB phase (see section 5.1). S stars or visual carbon
stars which have very thin dust shells are believed to be products of the
chemical transition from O to C in the AGB phase. The objects which become S
stars or visual carbon stars nearly at the end of the AGB phase are likely to
have very thin dust (silicate or AMC) shells at the beginning of the post-AGB
phase. These objects with with higher $v_{exp}$ could be the post-AGB stars or
PNe in the upper-left region (VI and LI) on the IRAS 2CD.

\section{Summary}

We have presented various IR 2CDs for AGB stars, post-AGB stars, and PNe
using catalogs from the available literature for the sample of 4903 AGB stars
(3373 O-AGB; 1168 C-AGB; 362 S-type), 660 post-AGB stars (326 post-AGB; 334
pre-PNe), and 1510 PNe in our Galaxy. For each object in the catalog, we
cross-identify the $IRAS$, $AKARI$, $MSX$, and 2MASS counterparts. For the
large sample of stars, we have presented various IR 2CDs using the $IRAS$
PSC, $AKARI$ PSC, $MSX$ PSC, and NIR (2MASS data at $K_s$ band) data.

To find possible evolutionary tracks from AGB stars to PNe on the 2CDs, we
have made simple but reasonable assumptions on the evolution of the dust
shell and the central star in the post-AGB phase. We assume that AGB dust
formation (at $T_c$ = 1000 K) ceases when the post-AGB phase starts. The dust
shell detaches with the start of post-AGB phase because there is no more dust
formation. We have performed radiative transfer model calculation for
detached dust shells around evolving central stars in the post-AGB phase.

We have found that the theoretical dust shell model tracks using dust opacity
functions of amorphous silicate and amorphous carbon roughly coincide with
the densely populated observed points of AGB stars, post-AGB stars, and PNe
on various IR 2CDs. Even though some discrepancies are inevitable, we have
found that the end points of the theoretical post-AGB model tracks are
generally converged to the region of observed points of PNe on most 2CDs. The
discrepancies could be due to the limitation of the theoretical model used
for this work. The dust shell model did not consider some dust species (e.g.,
PAH) and gas-phase radiation processes. The spherically symmetric model would
not be applicable to a major portion of the observed objects which are highly
non-spherical.

In this work, comparison of the model results with the observations on the IR
2CDs have indicated that the duration of the theoretical post-AGB phase needs
to be rather short (within a thousand years) to match the observations. This
could be because relatively small mass stars are not observed as post-AGB stars
or PNe. The duration of the post-AGB phase for very low-mass stars can be as
long as 10$^4$-10$^5$ years (see section 4.1). A major portion of these very
low-mass stars may evolve to the white dwarf stage without ever becoming a PN.

We have discussed two sequences of the post-AGB evolution on the $IRAS$ 2CD,
which are designated as LI (blue group) and RI (red group). The red objects
are likely to undergo dust shell detachments at the normal AGB termination.
Most of the observed PNe are converged to the RI region as well as the
results of all theoretical dust shell models with moderate $v_{exp}$. On the
other hand, most of the blue objects in the LI region are likely to be
non-spherical systems (precursors of bipolar PNe or binary systems) to which
the theoretical dust shell models may not be applicable. However, we have
demonstrated that it is also possible that the blue post-AGB stars in the LI
region as well as in the upper-left region of the $IRAS$ 2CD have very thin
dust shells with higher $v_{exp}$ because they had thin dust shells at the
starts of their post-AGB phases. We have argued that an objects in its early
post-AGB phase may have a very thin dust shells if the chemical transition
from O to C due to a thermal pulse occurred in its late AGB phase.

\section*{Acknowledgments}

This research was supported by Basic Science Research Program through the
National Research Foundation of Korea (NRF) funded by the Ministry of
Science, ICT \& Future Planning (NRF-2013R1A1A2057841). This research has
made use of the SIMBAD database, operated at CDS, Strasbourg, France. This
research is based on observations with $AKARI$, a JAXA project with the
participation of ESA.

\end{document}